\documentclass[journal]{IEEEtran}

\def\BibTeX{{\rm B\kern-.05em{\sc i\kern-.025em b}\kern-.08em T\kern-.1667em\lower.7ex\hbox{E}\kern-.125emX}}
\usepackage[numbers]{natbib}
\usepackage{graphicx}

\usepackage{indentfirst}
\usepackage{array,multirow}
\usepackage{fancyhdr}
\usepackage{float}
\usepackage{booktabs}
\usepackage{here}
\usepackage{comment}
\usepackage{siunitx}
\usepackage{algorithmicx,algorithm,algpseudocode}
\usepackage{amsmath,amssymb,amsfonts,bm,gensymb}
\usepackage{fontawesome5}
\usepackage{caption}
\usepackage{setspace}
\usepackage{cite}
 
\usepackage{svg}
\usepackage{media9}
\newcolumntype{R}{>{\centering\arraybackslash}m{0.8cm}}
\usepackage[normalem]{ulem}
\useunder{\uline}{\ul}{}
\usepackage{url,hyperref,microtype}
\usepackage{CJKutf8}

\usepackage{subcaption}
\usepackage[switch]{lineno}

\usepackage{color}

\definecolor{C}{RGB}{64, 224, 208}
\definecolor{O}{RGB}{246, 170, 0}

\newcommand{\etal}{et~al.\ }

\newcommand{\ie}{i.\,e.,\ }

\begin{document}


\title{An eHMI Presenting Request-to-Intervene and Takeover Status of Level 3 Automated Vehicles to Support Surrounding Traffic Safety}

\author{
Hailong~Liu$^{*}$,~\IEEEmembership{Senior Member,~IEEE},\
Masaki~Kuge,~\IEEEmembership{Student Member,~IEEE,} \\ Toshihiro~Hiraoka,~\IEEEmembership{Member,~IEEE} and Takahiro~Wada,~\IEEEmembership{Member,~IEEE}
\thanks{Hailong Liu, Masaki~Kuge and Takahiro Wada are with Graduate School of Science and Technology, Nara Institute of Science and Technology, 8916-5 Takayama-cho, Ikoma, Nara, 630-0192, Japan. }
\thanks{Toshihiro~Hiraoka is with Japan Automobile Research Institute, 1-1-30, Shibadaimon, Minato-ku, Tokyo 105-0012, Japan. }
\thanks{*Corresponding author: Hailong Liu~:~{\tt\small liu.hailong@is.naist.jp}}
}

\markboth{Journal of \LaTeX\ Class Files,~Vol.~14, No.~8, August~2015}%
{Shell \MakeLowercase{\textit{et al.}}: Bare Demo of IEEEtran.cls for IEEE Journals}

\maketitle

\begin{abstract}

Level~3 automated vehicles (AVs) issue a request to intervene (RtI) when the automated driving system approaches its system limitations. 
Although this takeover transition is safety-critical, it is usually invisible to surrounding manually driven vehicle (MV) drivers. 
This study proposes an external human-machine interface (eHMI) called eHMI C+O that externalizes the RtI-related takeover status of a Level~3 AV using cyan and orange light bars. 
A driving-simulator experiment with 40 participants examined whether the proposed eHMI supports surrounding MV drivers during AV takeover scenarios.
The results showed that, compared with the ADS-status-only eHMI condition, which is similar to ``Automated Driving Marker Lights,'' and the no-eHMI condition, the proposed eHMI C+O significantly improved participants' understanding of the AV's driving intention, their prediction of its behavior, and their perceived sufficiency of the information presented by the AV.
It also reduced hesitation, increased confidence, and promoted earlier and larger increases in time headway after the RtI was issued.
In the AV accident scenario, eHMI C+O significantly reduced the odds of accident involvement for the following MV compared with the no-eHMI condition, corresponding to a 76.8\% reduction in accident odds.
Exploratory path analysis suggested that the safety benefit of the proposed eHMI C+O may be associated with improved situation awareness and earlier defensive driving responses. 
These findings indicate that externalizing RtI-related takeover status can help surrounding drivers better understand Level~3 AVs and respond more safely during safety-critical takeover transitions.
\end{abstract}


\IEEEpeerreviewmaketitle

\section{INTRODUCTION}

\IEEEPARstart{A}{u}Autonomous vehicles (AVs) equipped with SAE levels 1-3 automated driving systems (ADS)~\citep{sae2018taxonomy} have steadily increased over the last few years~\citep{shanmugam2024autonomous}. 
Particularly, Level 3 AVs have attracted significant attention due to their potential to reduce driver workload and improve traffic safety~\citep{9046805, shanmugam2024autonomous}.
In Level 3 automated vehicles (AVs), drivers are allowed to disengage from the driving task under specific operational design domain (ODD) conditions~\citep{sae2018taxonomy}. 
While Level 3 ADS are performing the driving task, drivers are allowed to engage in non-driving-related tasks (NDRTs), such as using mobile phones~\citep{Wintersberger2021,Nabil2024}, watching videos~\citep{Pan02012025}, or playing games~\citep{jiang2024playing}. 
When Level 3 ADS encounter conditions beyond their ODD or functional limits, a Request to Intervene (RtI) is issued to the driver. 
In such a situation, the driver needs to promptly and properly take over control of the vehicle to ensure driving safety. 
If the driver fails to take over promptly, an accident may occur, potentially involving surrounding vehicles~\citep{li2021influence}.

\subsection{Takeover Challenges of Drivers in Level 3 AVs}

Although Level 3 ADS allow drivers to disengage from the driving task and engage in preferred NDRTs, several studies have reported that NDRTs can lead to driver distraction when an RtI is issued, making it difficult for drivers to respond promptly~\citep{merat2014transition, zeeb2016take, lu2017much}. 
Such distraction can delay drivers' reactions by increasing cognitive load and reducing situational awareness, resulting in longer reorientation times and greater variability in takeover performance~\citep{zhang2019determinants}. 
Typically, the takeover process requires approximately 10--15 seconds, followed by an additional 25--30 seconds for drivers to stabilize vehicle control~\citep{merat2014transition}. 
Moreover, \citet{gold2016taking} reported that takeover response times further increase under complex traffic conditions, such as on multi-lane roads or in high-density traffic.

Alternatively, if drivers of Level 3 AVs do not correctly understand the ODD or system limitations of the ADS they are using, they may develop over-trust in the ADS~\citep{liu2019driving}, which can cause them to miss the optimal opportunity to respond to an RtI and potentially result in a failed takeover~\citep{matsuo2026educational}. 
Based on the aforementioned issues, if a driver fails to promptly and safely take over control from the ADS, the probability of accidents will substantially increase~\citep{merat2014transition, zeeb2016take, lu2017much, gold2016taking, liu2019driving, matsuo2026educational}.

These studies have mainly focused on the AV driver's takeover performance, such as response time, takeover quality, and post-takeover control stability. 
However, from the perspective of traffic safety in MV--AV mixed traffic, an RtI is not merely a driver--vehicle interaction event inside the AV. 
Rather, it also indicates that the AV has entered a safety-critical transition state, in which the ADS is approaching or has reached its functional limit and the subsequent safety of the vehicle depends on uncertain human takeover performance. 
While this transition is known to the AV and its driver, it is largely invisible to surrounding road users, especially drivers of surrounding manually driven vehicles (MVs).

\subsection{Potential Risks of AV Takeover to Surrounding Vehicles}

\citet{alms2023control} showed that takeover events of Level 3 AV can trigger braking disturbances, gap fluctuations, and string-instability effects in mixed traffic, thereby introducing additional safety risks to surrounding vehicles, especially under high traffic demand and increasing AV penetration. 
We considered that if drivers of Level 3 AVs fail to promptly take control during an RtI, it not only endangers their own safety but may also pose risks to surrounding vehicles. 
Such delayed responses could trigger a chain reaction that compromises the safety of nearby traffic. 
For example, \citet{jia2024drivers} found that drivers of Level 3 AVs often neglected vehicles at their sides and rear during takeover transitions, and this lack of awareness was associated with a higher risk of collisions.

From the perspective of surrounding drivers, the risk introduced by RtI is compound. 
First, the issuance of RtI itself implies that the AV is encountering a scenario in which its automated driving capability is no longer fully reliable. 
Second, whether this risk escalates further depends on whether the human driver inside the AV can successfully resume control. 
As a result, surrounding drivers face uncertainty not only about the AV's current state but also about its near-future behavior. 
This uncertainty may make the AV's motion appear less predictable and may reduce the time available for surrounding drivers to prepare evasive actions, thereby increasing the likelihood of secondary collision involvement.

However, despite the safety risks posed by AV takeover to surrounding vehicles, existing studies have not explicitly considered RtI as a safety-critical transition state that is externally invisible to surrounding drivers. In addition, existing external communication methods do not explicitly inform surrounding drivers whether an RtI has been issued or what the AV driver's takeover status is after RtI. Therefore, it remains unclear how externalizing such information may influence surrounding drivers' cognition, safety-margin regulation, and collision involvement.

\subsection{Purpose and Contributions}

In this study, we propose an RtI-related eHMI that externalizes the safety-critical transition state of a Level 3 AV to surrounding drivers. Unlike conventional eHMIs, which primarily indicate the current driving mode, the proposed interface provides explicit information about both the activation of automated driving and the issuance of an RtI, thereby enabling surrounding drivers to recognize that the AV has entered a potentially hazardous transition phase.

We hypothesize that externalizing the RtI state in this manner can reduce surrounding drivers' uncertainty about the AV's condition, improve their situation awareness, and promote earlier and more appropriate safety-margin regulation, such as increasing time headway (THW).
Importantly, while the proposed eHMI is not intended to prevent the AV crash itself in the event of takeover failure, it may mitigate the risk of secondary collision involvement by enabling surrounding drivers to prepare in advance and respond more effectively.

To test this hypothesis, we conducted a driving simulator experiment in which participants drove an MV following a Level 3 AV under different eHMI conditions and takeover scenarios. 
We evaluated (1) subjective cognition related to the understanding and prediction of the AV's behavior, (2) behavioral responses in terms of safety-margin regulation, and (3) the occurrence of collision involvement when the AV driver failed to take over.

The contributions of this study are threefold. 
First, this study conceptualizes RtI as a safety-critical transition state that is internally known to the AV but difficult for surrounding drivers to perceive externally.
Second, this study proposes an eHMI that externalizes not only the ADS status and the issuance of RtI but also the AV driver's takeover status after RtI. 
Third, this study provides an exploratory investigation of how such information affects surrounding drivers' cognition, safety-margin regulation, and collision involvement in AV-initiated crash scenarios.

\section{RELATED WORK}

\subsection{Policies to Show the Potential Risks of AVs}

Currently, governments and manufacturers in many countries recognize the technical limitations and potential risks of AVs. 
Accordingly, some countries have introduced policies requiring vehicles equipped with Level 3--5 ADSs to be explicitly identifiable as AVs by other road users (ORUs).

In Japan, the Ministry of Land, Infrastructure, Transport and Tourism (MLIT) has proposed the use of an ``\textit{Automated Drive Sticker}''~\citep{MLITsticker} to make vehicles equipped with Level 3--5 ADSs readily identifiable from the outside.

In the United States, SAE Recommended Practice J3134~\citep{SAEJ3134_2019} defines the technical characteristics of \textit{ADS marker lamps}, which illuminate in cyan when the ADS is active. 
The motivation behind these developments is to enhance surrounding road users' confidence in automated vehicles. 
Building upon SAE J3134, Mercedes-Benz Group AG. proposed a light-based eHMI called ``\textit{Automated Driving Marker Lights}''~\citep{Mercedes-turqoise} in 2024. 
It is designed to externally indicate the operational status of the ADS and to distinguish between automated and manual driving modes. 
According to publicly available information, this lighting system has been approved for use in California and Nevada, United States, and has already been implemented on vehicles.

Currently (April 2026), in China, although no official regulations specifically governing dedicated signal lights or markers for autonomous vehicles have yet been formally enacted, several Chinese automakers have already introduced cyan ADS marker lamps on commercially available vehicles~\citep{gibbs2025turquoise}. 
At present, the operational logic of these automated driving lights is broadly similar to that used in the United States. 
Specifically, the ADS marker lamps are designed to illuminate automatically when the ADS is active and to turn off automatically when the system is inactive.

While the policies and methods described above help ORUs recognize whether a vehicle is equipped with an ADS and whether it is currently operating in automated driving mode, it remains difficult for them to determine whether a Level 3 AV has issued an RtI or what the AV driver's takeover status is after the RtI. 
Consequently, ORUs may be unable to take timely and appropriate preventive actions.

\subsection{External Human--Machine Interfaces for Communication of AV with surrounding MV drivers}

An external Human--Machine Interface (eHMI) enables AVs to convey their status and communicate with ORUs, such as pedestrians and other vehicles. 
Indicator lights, brake lights, and horns can also be regarded as conventional eHMI methods. 
To enhance road safety and efficiency, recent studies~\citep{Clercq2019eHMI, dey2021towards, liu2021importance, li2021autonomous, li2022hmi, lee2022learning, li2023av, liu2025pre} have proposed using light bars and displays as innovative eHMIs to provide ORUs with more explicit information, such as yielding intentions. 
The vast majority of eHMI research has focused on communication between AVs and pedestrians~\citep{Clercq2019eHMI, dey2021towards, liu2021importance, lee2022learning,liu2025pre}.

\begin{figure*}[!bt]
\footnotesize
    \centering
    \includegraphics[width=1\linewidth]{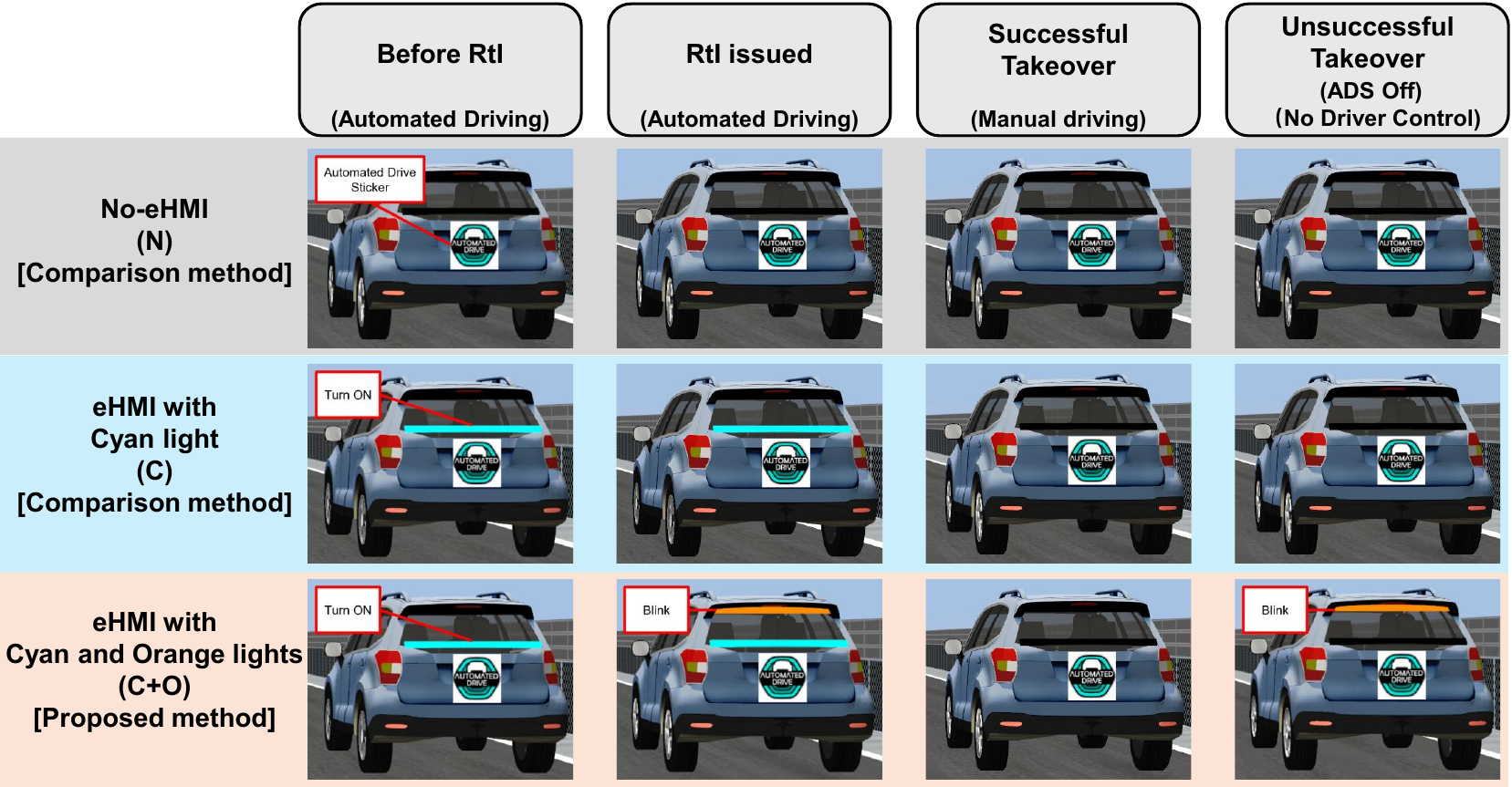}
    \caption{The proposed eHMI and the other eHMIs used for comparison in the experiment.}
    \label{fig:eHMI}
\end{figure*}

At present, only a few studies have addressed communication between AVs and drivers of surrounding MVs via eHMIs. 
For example, \citet{stange2022manual} reported that an eHMI displaying ADS status, whose function is essentially equivalent to automated driving marker lights, helped surrounding drivers identify automated vehicles, but did not significantly affect their subjective evaluations or driving behavior. 
This suggests that merely communicating the current ADS status via the eHMI may be insufficient to support surrounding manual drivers in mixed traffic environments with AVs.
Similarly, for ADS status eHMI of a Level 4 autonomous truck, \citet{gwak2025encounter} reported that it could enhance the AV’s conspicuity and facilitate surrounding drivers' early attention at highway junctions.

In contrast, some eHMIs have been designed to communicate the AV’s driving intention, and such eHMIs have been shown to facilitate communication between AVs and MV drivers during right-of-way negotiation at bottleneck roads~\citep{rettenmaier2020after, li2023_AutoUI} and T-junctions~\citep{avsar2021efficient,lingam2024ehmi}.
Furthermore, to address multi-vehicle interactions, \citet{li2022hmi, li2023av} proposed combining an eHMI with an internal HMI based on vehicle-to-vehicle communication to facilitate targeted communication of driving intention to a specific MV in bottleneck-road scenarios.
For Level 4 AVs, \citet{schindler2020communicating} and \citet{hub2023promoting} proposed using eHMIs to communicate warning messages to following MVs when Level 4 AVs perform minimal risk maneuvers (MRMs) in traffic scenarios near the boundaries of their ODDs.

To the best of our knowledge, the use of eHMIs on Level 3 AVs to deliver early warning information to ORUs when an RtI is issued remains underexplored. 
Moreover, existing studies have rarely considered RtI as a safety-critical transition state that is externally difficult for surrounding drivers to perceive, nor have they explicitly investigated how communicating the AV driver's takeover status after RtI may affect surrounding drivers' cognition, safety-margin regulation, and collision involvement.
To address this gap, our preliminary study~\citep{IV2024_kuge} proposed an eHMI that displayed the ADS status and the AV driver's takeover outcome after an RtI. 
By presenting these two types of information, surrounding MV drivers could become aware in advance of potential risks caused by delayed or unsuccessful takeover by the AV driver. 
However, in that preliminary study~\citep{IV2024_kuge}, only 12 participants were recruited as MV drivers. 
Although no significant effects on MV drivers' driving behavior were observed, participants reported a greater willingness to have AVs equipped with the proposed eHMI driving around them in their daily lives. 
Therefore, in the present study, the number of participants was increased to 40, and the effects of AVs equipped with different eHMIs on following MV drivers' psychological states and driving behavior during the takeover process were further and systematically analyzed across multiple driving scenarios.

\section{Proposed Method}
\label{sec:eHMI_C+O}

This study conceptualizes RtI as a safety-critical transition state and aims to externalize this otherwise hidden state to surrounding drivers through eHMI design. 
From the perspective of surrounding MV drivers, this study considers the following information to be important for helping them become aware of potential risks associated with Level~3 AVs during the takeover process:
\begin{itemize}
    \item[(1)] the operational status of the ADS;
    \item[(2)] whether an RtI has been issued;
    \item[(3)] the AV driver's takeover status after RtI.
\end{itemize}
To explicitly convey the above information from a Level~3 AV to surrounding drivers, we propose an eHMI consisting of two light bars in cyan and orange, mounted on the rear of the AV, as shown in the bottom row of Fig.~\ref{fig:eHMI}. 
The light bars were designed with a horizontally elongated shape to enhance visibility and to avoid confusion with brake lights or turn signals.

The cyan light bar (Hex color: \colorbox{C}{\#40E0D0}) was designed in accordance with SAE~J3134~\citep{saej3134}. 
It is installed along the lower edge of the rear window of the AV to indicate the operational status of the ADS. 
Specifically, the cyan light bar is activated when the ADS is in use and remains illuminated until the ADS is turned off.

Instead of using the conventional hazard warning lights after an RtI is issued, this study introduced a dedicated orange light bar to represent the RtI state. 
The main reason is that hazard warning lights are conventionally associated with emergency stopping, vehicle breakdowns, or a vehicle that is already in a hazardous condition, while a Level~3 AV that has issued an RtI still be operating under active ADS control for a limited period.
Therefore, reusing hazard lights may cause surrounding drivers to misunderstand the AV's actual state. 
The color and lighting pattern of the orange light bar (Hex color: \colorbox{O}{\#F6AA00}), which serves as a warning signal, follow the ISO safety color standard~\citep{international1984international} and operate at a flashing frequency of 1~Hz with a 50\% duty cycle (0.5~s on, 0.5~s off). 
It is installed along the upper edge of the rear window to indicate that an RtI has been issued. 
When the AV issues an RtI to its driver, the orange light bar is activated and starts blinking, indicating a potential driving risk posed by the AV to surrounding vehicles.

It is important to note that the ADS issues an RtI to its driver a few seconds before reaching its system limits. 
During this period, not only does the orange light bar flash, but the cyan light bar also remains illuminated, because the ADS remains active after the RtI is issued.
After the RtI is issued and the AV driver completes the takeover, the orange light bar is extinguished to indicate successful takeover by the AV driver, and the cyan light bar is also extinguished to denote a transition to manual driving mode.
In contrast, if the AV driver does not take over the driving task after the ADS is forcibly deactivated due to the vehicle exceeding its system limits, \ie no driver control, the cyan light bar is turned off to indicate that the ADS is inactive, while the orange light bar continues to flash, signifying that the vehicle has not been taken over by the AV driver.

\section{EXPERIMENT}

This study conducted a driving simulator (DS) based experiment with a within-subject design to examine whether the proposed eHMI can reduce collision risks in mixed traffic by presenting the AV's takeover status. 
Participants were assigned the role of drivers of manually driven vehicles (MVs) following the AV. 
The experiment included three eHMI conditions and two AV takeover outcomes after RtI (\ie successful takeover and unsuccessful takeover), implemented across two road contexts.
This design was intended to prevent participants from interpreting every RtI event as a deterministic precursor to a crash and from adopting uniformly defensive driving strategies in all RtI trials. 
The experiment received approval from the Research Ethics Committee of Nara Institute of Science and Technology (No. 2023-I-11).

\subsection{Participants}

An \textit{a priori} power analysis was conducted using \textit{G*Power} (Version 3.1.9.7~\citep{GPower}) for a repeated-measures ANOVA. 
Assuming $f = 0.25$, $\alpha = 0.05$, and $1-\beta = 0.80$, the required sample size was estimated to be 28 for the full $2 \times 3$ within-subject design. 
Accordingly, 40 participants aged 22--24 years took part in the experiment as drivers of a manually driven vehicle. 
All participants held a valid Japanese driving license. 
Before the experiment, informed consent was obtained from each participant. 
Each participant took approximately one hour to complete the experiment. 
After completing the experiment, each participant received 1,000 Japanese yen as compensation.

\subsection{Driving Simulator}

\begin{figure}[!t]
    \centering
    \includegraphics[width=1\linewidth]{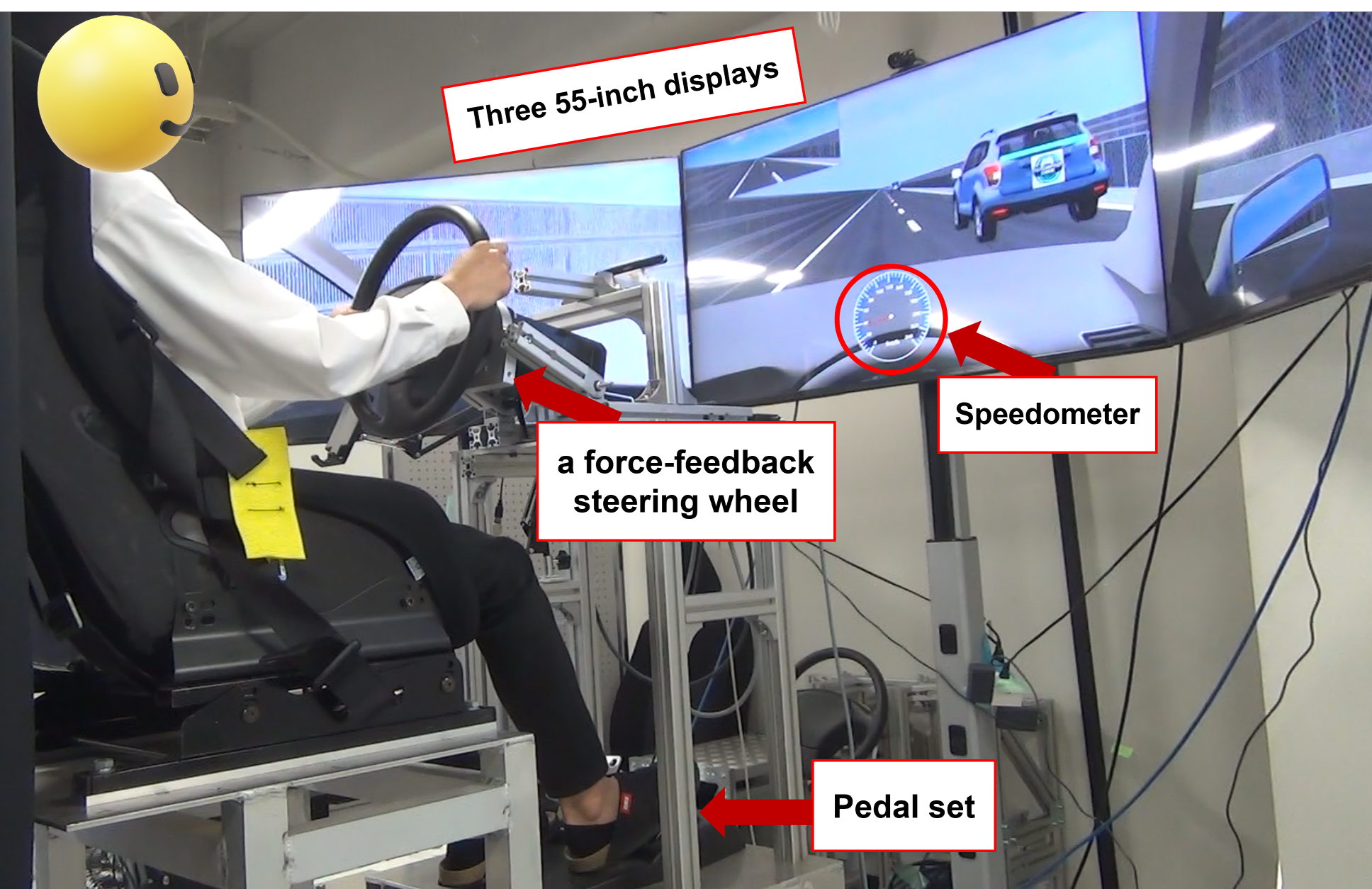}
    \caption{The diving simulator used in the experiment.}
    \label{fig:DS}
    \vspace{-3mm}
\end{figure}

A driving simulator (DS) based on UC-win/Road Ver.~16 (FORUM 8 Co., Ltd.) was used in this experiment (see Fig.~\ref{fig:DS}). 
The driving simulation environment was presented to the participants using three 55-inch displays, each with a resolution of 1920 $\times$ 1080 pixels. 
Participants operated a manually driven automatic-transmission vehicle in the simulated driving environment using a Logitech G29 pedal set and a SENSO Wheel SD-LC force-feedback steering wheel.


\subsection{Successful and Unsuccessful AV Takeover Scenarios}

In this study, the main experimental factors were eHMI condition and AV takeover outcome, \ie whether the AV driver successfully resumed control after the RtI was issued. 
Accordingly, an \textit{AV successful takeover} scenario and an \textit{AV unsuccessful takeover} scenario were designed for the experiment.

\begin{figure}[!t]
    \centering
    \subfloat[AV is successfully taken over on a curve road\label{fig:curve-S}]{
        \includegraphics[width=\linewidth]{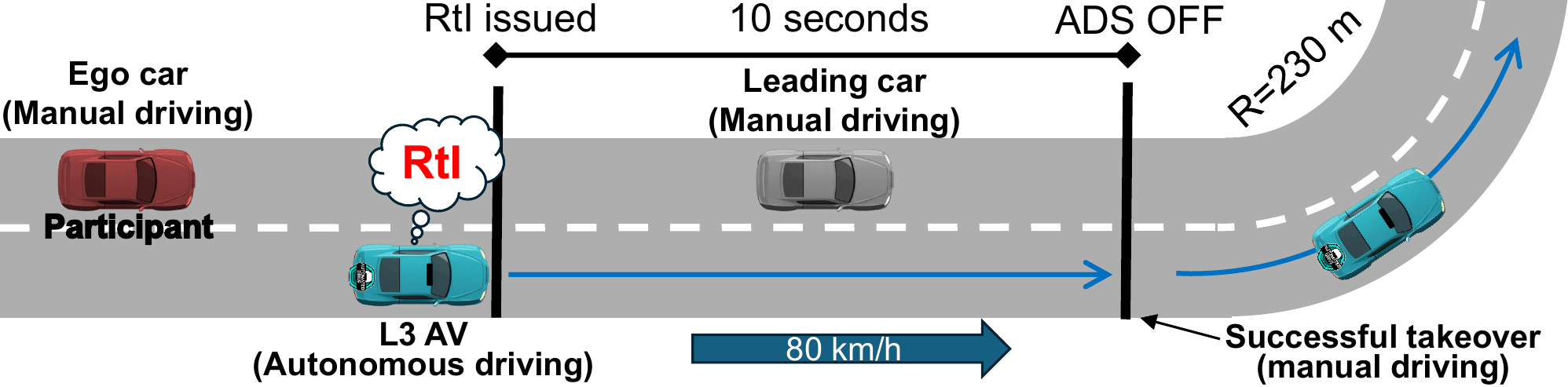}
    }
    \\
    \vspace{2mm}
    \subfloat[AV is unsuccessfully taken over on a curve road\label{fig:curve-A}]{
        \includegraphics[width=\linewidth]{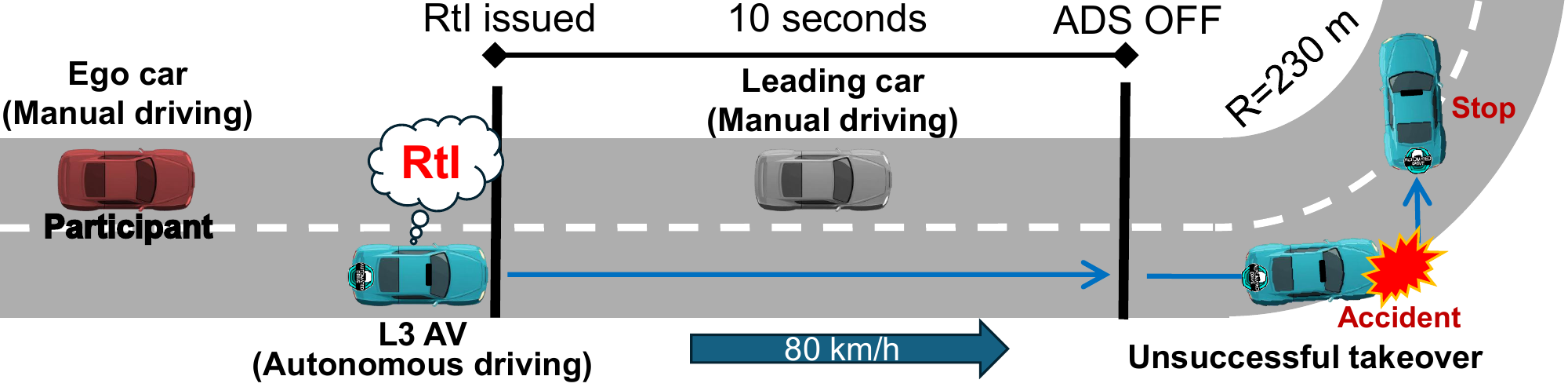}
    }
    \\
        \vspace{2mm}
    \subfloat[AV is successfully taken over on a diverging road\label{fig:exit-S}]{
        \includegraphics[width=\linewidth]{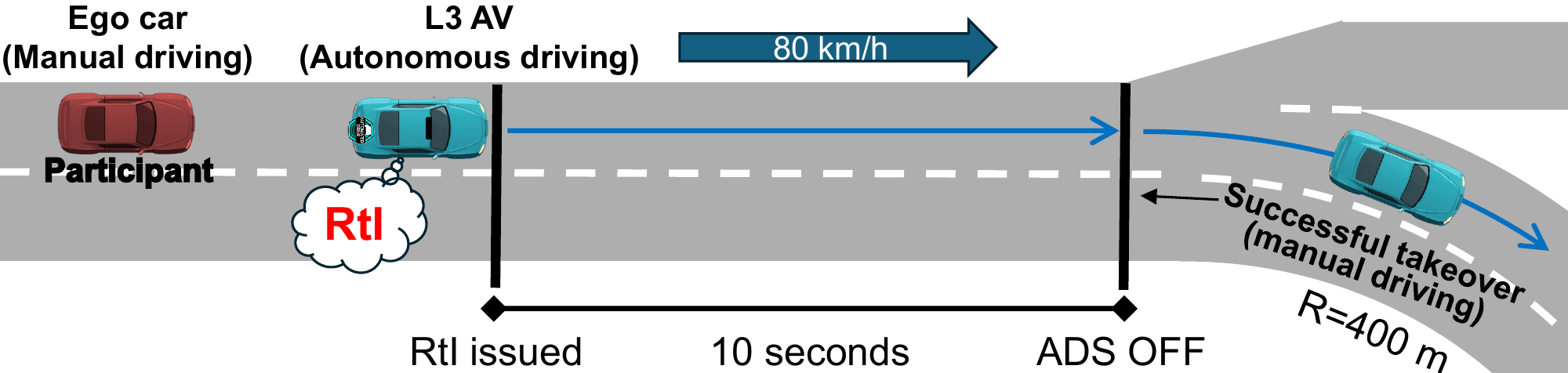}
    }
    \\
        \vspace{2mm}
    \subfloat[AV is unsuccessfully taken over on a diverging road\label{fig:exit-A}]{
        \includegraphics[width=\linewidth]{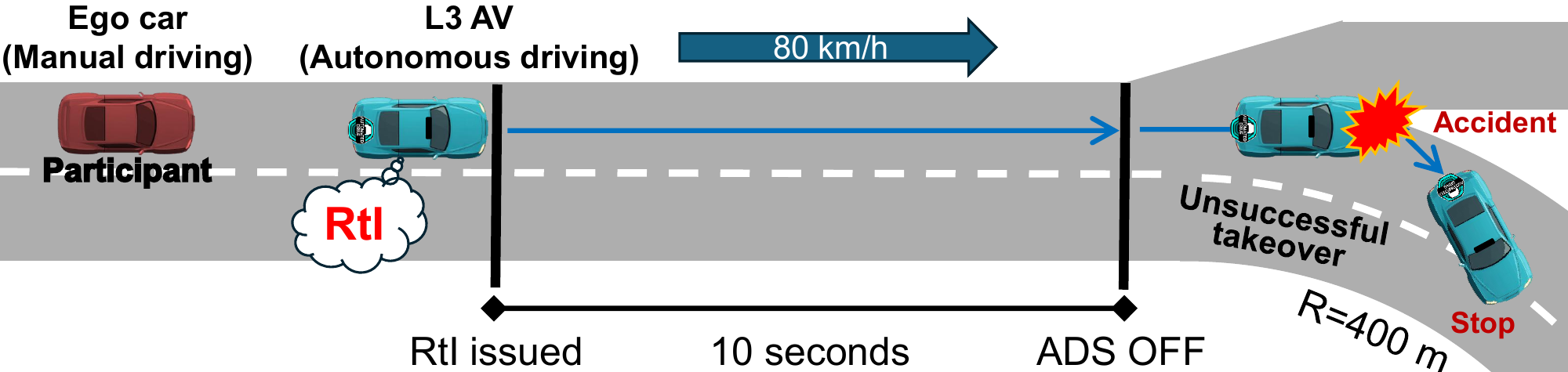}
    }        
    \caption{Driving simulator scenarios used in the experiment. The red car represents the ego MV driving by the participants, the blue car represents the AV, and the gray car represents another MV. The blue lines denote the trajectories of the AV.}
    \label{fig:courses}
\end{figure}

To avoid making the driving scenarios overly monotonous and to provide participants with more realistic encounter situations, two planned RtI situations were assumed for the AV, and corresponding AV successful and unsuccessful takeover scenarios were designed.
As shown in Fig.~\ref{fig:courses}, these two RtI situations were respectively implemented in two highway driving contexts, \ie \textit{Curve road} and \textit{Diverging road}.

On the \textit{Curve road}, as shown in Fig.~\ref{fig:curve-S} or ~\ref{fig:curve-A}, a Level~3 AV drove at 80~km/h in the right lane ahead of the MV driven by the participant. 
After 1~km of straight driving, the AV was assumed to encounter a system limitation such that it could not automatically negotiate a curve with a radius of 230~m at 80~km/h. 
This assumption was based on Article 15 (Curve Radius) of the Road Structure Ordinance of Japan~\citep{JP_RSO} and \citet{matsuo2026educational}.
This was based on the assumption that sharp curves would hinder the AV's ability to detect lane markings and traffic conditions far enough ahead in a timely manner.
Therefore, the AV issued an RtI 10~s before entering the curve ($R=230$), and the ADS was turned off when the AV entered the curve. 
In the \textit{\textbf{AV successful takeover}} scenario (see Fig.~\ref{fig:curve-S}), the AV driver resumed control at 10~s after the RtI was issued and safely drove through the curve.
In the \textit{\textbf{AV unsuccessful takeover}} scenario (see Fig.~\ref{fig:curve-A}), the AV driver did not take over within 10~s after the RtI was issued. 
As a result, the AV collided with the wall in the curve and rebounded toward the center of the lane in which the participant was driving.

On the \textit{Diverging road}, an RtI scenario was designed based on an actual AV accident~\citep{tesla-accident}.
Specifically, a Level~3 AV drove at 80~km/h ahead of the MV driven by the participant. 
After 1~km of straight driving, the AV approached a highway diverging area that it was designed to pass through rather than exit (see Figs.~\ref{fig:exit-S} and~\ref{fig:exit-A}). 
Because of the complexity of lane markings in this area, the diverging section was assumed to be beyond the AV's system capability. 
Therefore, the AV issued an RtI 10~s before reaching the diverging area. 
In the \textit{\textbf{AV successful takeover}} scenario (see Fig.~\ref{fig:exit-S}), the AV driver resumed control within 10~s after the RtI was issued and safely drove through the diverging area. 
In the \textit{\textbf{AV unsuccessful takeover}} scenario (see Fig.~\ref{fig:exit-A}), the AV driver did not take over within 10~s after the RtI was issued. 
As a result, the AV collided with the merge divider and rebounded toward the center of the lane in which the participant was driving, reproducing the actual AV accident~\citep{tesla-accident}.

\subsection{eHMI Conditions}

To validate the effectiveness of the proposed method, \ie \textit{eHMI with Cyan and Orange lights} (eHMI C+O), introduced in section~\ref{sec:eHMI_C+O}, two eHMI conditions, \textit{no-eHMI} (eHMI N) and \textit{eHMI with cyan lights} (eHMI C), were developed as contrasting methods.

In the eHMI N condition, as shown in the top row of Fig.~\ref{fig:eHMI}, the AV is not equipped with any eHMI light bar.
The AV does not provide any explicit cues to the outside during the transition from automated driving to manual driving.

In the eHMI C condition, only the cyan light bar is equipped on the AV (see the middle row of Fig.~\ref{fig:eHMI}), simulating the use case of the \textit{Automated Driving Marker Lights}.
Under this condition, eHMI C only informs surrounding drivers of the ADS status and does not provide information about the takeover status during the takeover process.

Additionally, to assist participants in accurately distinguishing between AV and MV driving in their surroundings, the ``\textit{Automated Drive Sticker}''~\citep{MLITsticker} was affixed to the rear of the AV (see Fig.~\ref{fig:eHMI}) under all conditions.

\subsection{Experiment Procedure}

First, the participants were informed about the purpose of the experiment and received instructions on how to use the DS.
After confirming their understanding and willingness to participate, participants were asked to complete a written consent form.

Subsequently, participants were informed that they would drive an MV on a highway in mixed traffic with other MVs and AVs.
They were then informed that the AVs were equipped with Level~3 ADS, and basic knowledge about Level~3 AVs and RtI was explained.
All AVs had an ``\textit{Automated Drive Sticker}'' affixed to their rear. 
Among the three types of Level~3 AVs, one had no-eHMI, while the other two were equipped with different eHMI designs.
Then, they were instructed the meanings of the information conveyed by the two types of eHMI, \ie \textit{eHMI with cyan light} and \textit{eHMI with cyan and orange lights}.
Regarding the driving task, participants were informed that they would use the DS to drive an MV safely at 80~km/h on a highway. 
They were also told that various vehicles, including MVs and one of the three types of Level~3 AVs, would appear randomly around them during the experiment.
As a cover story, participants were informed that another participant was acting as the Level~3 AV driver in a different room using another driving simulator. They were told that, when the AV encountered a system limitation, an RtI would be issued to that participant as the AV driver. Accordingly, whether the AV was successfully taken over depended on that driver's response to the RtI.

After the pre-instruction phase, participants practiced driving the MV using the DS. 
They then performed 12 trials, consisting of three eHMI conditions and two AV takeover conditions on two driving courses. 
The trial order was randomized for each participant.

After each trial, participants were required to answer a post-trial questionnaire (see section~\ref{sec:PTQ}).
After completing the 12 trials, participants were also asked to answer a post-experiment questionnaire (see section~\ref{sec:PEQ}).

\subsection{Measurements and Data Analysis}

\vspace{2mm}
\subsubsection{\textbf{Post-trial Questionnaire}}
\label{sec:PTQ}

Referring to the cognition--decision--behavior model proposed by Liu \etal\citep{liu2023implicit}, six questions were used to evaluate participants' subjective responses in each trial. The questions were categorized into three groups as follows:

\textit{(1) Cognition-related subjective evaluations}
\begin{itemize}
\small{
    \item[Q1]: Was it easy to understand the driving intention of the AV?
    \item[Q2]: Was it easy to predict the driving behaviors of the AV?
    \item[Q3]: Was the information presented by the AV sufficient for you?
}
\end{itemize}

\textit{(2) Risk-related subjective evaluation}
\begin{itemize}
\small{
    \item[Q4]: Did you feel that the AV's behavior was dangerous?
    }
\end{itemize}

\textit{(3) Response-related subjective evaluations}
\begin{itemize}
\small{
    \item[Q5]: Did you confidently handle the driving operation in this trial?
    \item[Q6]: Did you hesitate in responding to the AV's behaviors?
}
\end{itemize}
Each question was answered using a 5-point Likert scale (1: completely disagree to 5: completely agree).

The ART ANOVAs with mixed-effects models~\citep{wobbrock2011aligned} were used to examine the effects of eHMI, takeover outcome of the AV (\texttt{Takeover}), and their interaction on each post-trial questionnaire item while accounting for repeated measurements within participants.
The model formula was 
\begin{equation}
\texttt{DV $\sim$ eHMI * Takeover + (1|ID)},\nonumber
\end{equation}
where \texttt{DV} denotes the dependent variable, corresponding to each questionnaire item from Q1 to Q6.
Significant main or interaction effects were followed by post hoc pairwise comparisons using ART-C~\citep{elkin2021aligned}, with FDR-adjusted p-values.

\vspace{2mm}
\subsubsection{\textbf{Overall Evaluation via Post-experiment Questionnaire}}
~\label{sec:PEQ}
At the end of all 12 trials, participants were asked to respond to the following two questions using a ranking method (1st place: most likely $\sim$ 3rd place: least likely):
\begin{itemize}
\small{
    \item[E1]: Please rank the eHMIs according to how likely their AVs are to involve other vehicles in traffic accidents during the takeover phase.
    \item[E2]: Please rank the eHMIs according to your preference for having their AVs driving around your vehicle in daily life.
}
\end{itemize}

Friedman tests were used to examine differences among eHMI conditions for each question, followed by Wilcoxon signed-rank tests with FDR correction for post hoc comparisons.

\vspace{2mm}

\subsubsection{\textbf{Time Headway Change and Its Area Under the Curve}}

To evaluate the influence of different eHMI conditions on the safety margin between the ego MV and the AV, the speed of the ego MV and the inter-vehicle distance between the ego MV and the AV were measured from the moment the AV issued the RtI. 
These measures were used to compute the time headway (THW) as $\mathrm{THW}(t) = \frac{D(t)}{V_{\mathrm{MV}}(t)}$, where $D(t)$ denotes the inter-vehicle distance from the front bumper of the ego MV to the rear end of the AV at time $t$, and $V_{\mathrm{MV}}(t)$ denotes the speed of the ego MV at time $t$.
Because the ego MV was manually driven by the participant in every trial, its cruising speed varied across trials.

To statistically examine how the ego MV driver changed the following distance after the AV issued the RtI under different eHMI conditions, the area under the curve (AUC) of THW change relative to the THW at RtI issuance was calculated for each trial within a time window of $T$ seconds as
\begin{equation}
\small
\texttt{AUC}^{\texttt{THW}}_{T}=\int_{t_{\mathrm{RtI}}}^{t_{\mathrm{RtI}}+T} \left(\mathrm{THW}(t)-\mathrm{THW}(t_{\mathrm{RtI}})\right)\,dt \nonumber
\end{equation}
where $dt=0.2$~s.

To examine when after the RtI the eHMI began to produce differences in THW, $T \in \{1, 2, 3, 4, 5, 10, 12\}$ s was considered.
Specifically, $\texttt{AUC}^{\texttt{THW}}_{1s}$ to $\texttt{AUC}^{\texttt{THW}}_{5s}$ were used to capture the temporal emergence of the ego MV driver's early driving responses immediately after the RtI.
$\texttt{AUC}^{\texttt{THW}}_{10s}$ was used to represent driving behavior during the RtI period, because the AV driver was required to take over within 10 s after the RtI was issued. 
In contrast, $\texttt{AUC}^{\texttt{THW}}_{12s}$ also included the period after the AV's takeover outcome became observable through the eHMI, and thus could reflect the influence of the takeover outcome on the ego MV driver's behavior. 
A larger $\texttt{AUC}^{\texttt{THW}}_{T}$ indicates that the ego MV driver increased the time headway earlier and/or more persistently after the RtI, reflecting more defensive driving behavior.

Because the AV driver was programmed to successfully take over at 10 s after the RtI (or fail to take over), THW within the 0–10~s interval was independent of the takeover condition. 
Therefore, for the AUC of THW changes within 5 s and 10 s after RtI, ART ANOVAs based on mixed-effects models~\citep{wobbrock2011aligned} were conducted to examine the effects of eHMI, takeover outcome of the AV (Takeover), and their interaction, while accounting for repeated measurements within participants via model formulas:
\begin{equation}
\texttt{$\texttt{AUC}^{\texttt{THW}}_{T\in\{1,2,3,4,5,10\}s}$ $\sim$ eHMI + (1|ID)}\\ \nonumber
\end{equation}
For the AUC within 12 s after RtI, which includes the AV takeover phase, ART ANOVAs~\citep{wobbrock2011aligned} were conducted to examine the effects of eHMI, Course, Takeover condition, and their interactions via:
\begin{equation}
\texttt{$\texttt{AUC}^{\texttt{THW}}_{12s}$ $\sim$ eHMI * Takeover + (1|ID)} \nonumber
\end{equation}
When significant main or interaction effects were detected, post hoc pairwise comparisons were performed using ART-C~\citep{elkin2021aligned}, with p-values adjusted via false discovery rate (FDR) method.

\vspace{2mm}
\subsubsection{\textbf{MV accident occurrence in the AV accident scenario}}

To examine whether participants driving the ego MV were more likely to be involved in an accident under different eHMI conditions following the AV's failed response to the RtI, accident occurrence was analyzed only in the AV accident scenario. 
The outcome of each trial was coded as a binary variable (1 = accident, 0 = no accident).

Fisher’s exact tests were first conducted to examine whether accident occurrence differed among the three eHMI conditions. 
Accordingly, to examine the overall effect of eHMI condition on MV accident occurrence, the primary model, referred to as Model 1, was specified as a binomial generalized linear mixed-effects model, with ego MV accident occurrence as the binary outcome and participant ID as a random intercept:
\begin{equation}
\texttt{MV$_{Accident}$ $\sim$ eHMI + (1|ID)}.\label{eq:main_model}
\end{equation}

\subsection{Exploratory Underlying Mechanisms Analysis of the eHMI Effect on MV Accident}

An additional exploratory analysis was conducted to examine the candidate dependency structure linking eHMI condition to MV accident occurrence through subjective evaluations (Q1 to Q6) and behavioral response variables derived from THW AUC measures. 

\vspace{2mm}
\subsubsection{\textbf{Gaussian Bayesian Network for Candidate Dependency Structure Estimation}}

An exploratory Gaussian Bayesian network (GBN) was learned to identify a candidate dependency structure among the variables.
It was learned in \texttt{R 4.5.3} using the \textit{tabu} search algorithm implemented in the \texttt{bnlearn} package, with a Gaussian BIC score~\citep{scutari2010learning}. 
This approach provides an approximate representation of the dependency structure and is intended for exploratory use.

To ensure consistency with the experimental design and temporal ordering, prior constraints were imposed during structure learning.
Specifically, MV accident occurrence was treated as a terminal outcome and was not allowed to have outgoing edges.
eHMI condition was treated as an exogenous manipulated variable and was not allowed to have incoming edges.
Edges from longer-window behavioral variables to shorter-window variables were prohibited (e.g., from $\texttt{AUC}^{\texttt{THW}}_{12s}$ to $\texttt{AUC}^{\texttt{THW}}_{5s}$ or $\texttt{AUC}^{\texttt{THW}}_{4s}$).

Note that, in this part, all variables were numerically encoded.
The eHMI condition and MV accident occurrence were integer-encoded, while subjective evaluation scores (Q1 to Q6) and THW AUC measures were treated as continuous variables.

\vspace{2mm}

\subsubsection{\textbf{Piecewise Structural Equation Modeling for Path-Based Examination}}

Based on the candidate structure obtained from the GBN analysis, an exploratory piecewise structural equation model (pSEM)~\citep{lefcheck2016piecewisesem} was constructed.
This approach was adopted because the variables involved in the exploratory mechanism analysis were mixed in type.
Specifically, the analysis included a categorical predictor (\ie eHMI condition), continuous variables (\ie Q1 to Q6 and the THW AUC measures), and a binary outcome variable (\ie MV accident occurrence).
In addition, the data had a repeated-measures structure, with multiple observations nested within participants.

In particular, pSEM allows different component component models to be specified according to the distributional characteristics of individual endogenous variables.
In this part, the eHMI condition was integer-encoded and MV accident occurrence was binary-encoded.
Other values were treated as continuous variables.
Therefore, linear mixed-effects models could be used for component models with continuous variables, whereas a binomial generalized linear mixed-effects model could be used for component model of the MV accident occurrence, making pSEM more suitable for the present data structure.

After constructing the pSEM based on the initial candidate model via the GBN, tests of directed separation were used to identify potentially missing edges. 
A significant independence claim suggested that the conditional independence implied by the current path structure was violated, indicating a possible omitted path between the corresponding variables. 
When such a dependency was also theoretically interpretable in the context of the experiment, the corresponding edge was added to the model. 
Fisher’s $C$ statistic was then used to assess whether the overall fit of the revised pSEM improved, and this process was repeated until the final model was determined.

\section{RESULTS}

\subsection{Post-trial Questionnaire}

\begin{figure}[!t]
    \centering
    \includegraphics[width=\linewidth]{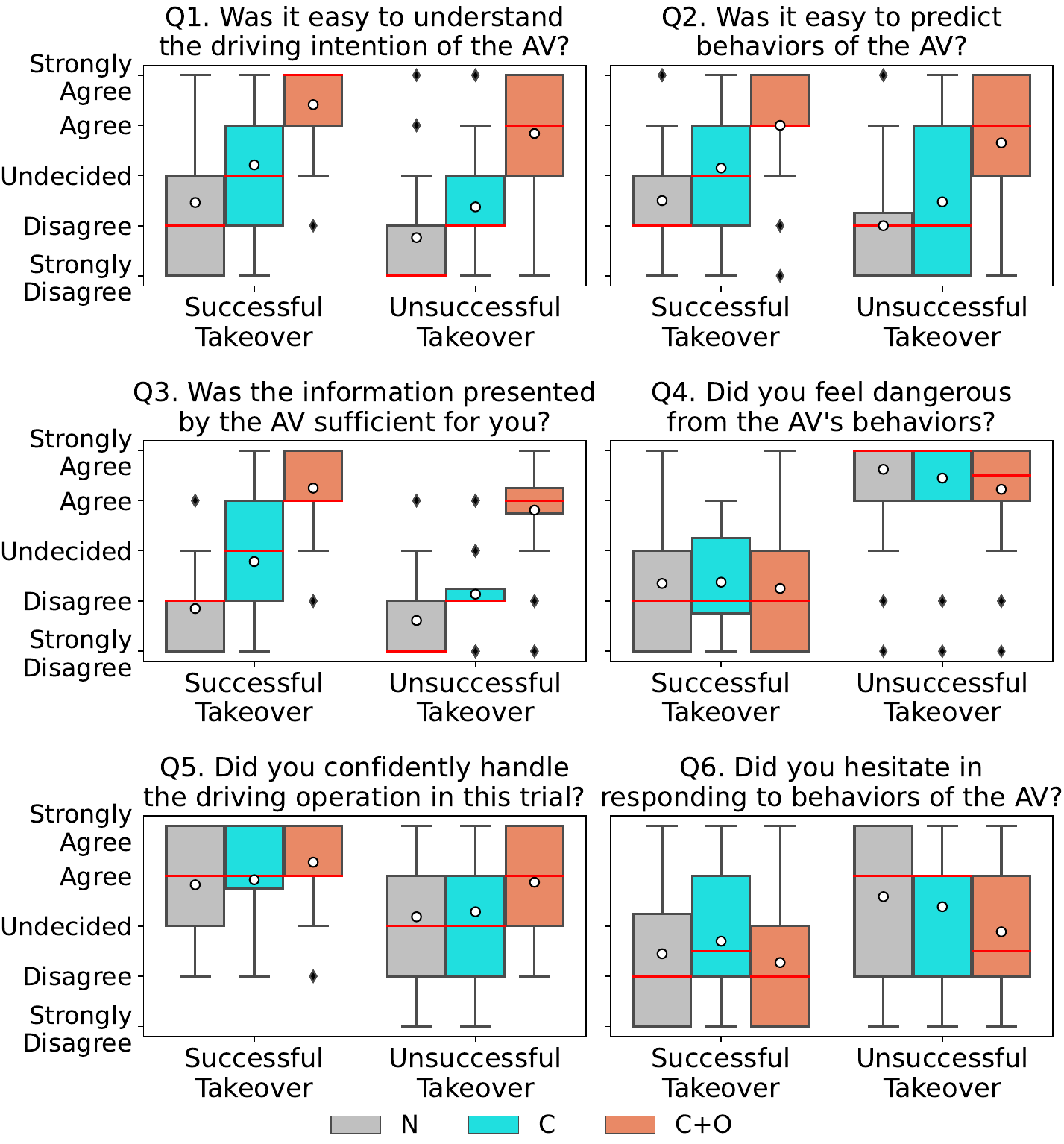}
    \caption{The results of the post-trial questionnaire.}
    \label{fig:PTQ}
\end{figure}

Table~\ref{tab:PTQ-ANOVA} summarizes the ART-based mixed-effects ANOVA results for the post-trial questionnaire items. 
The main effect of eHMI was significant for Q1, Q2, Q3, Q5, and Q6 (all $p<.001$), but not for Q4 ($p=.158$).
The main effect of takeover outcome was significant for all six items (all $p<.001$). 
No significant eHMI $\times$ takeover interaction was found for Q1--Q4 (all $p>.05$), while significant interactions were observed for Q5 ($p=.037$) and Q6 ($p=.021$).

\begin{table}[!t]
\setlength\tabcolsep{7pt}
\footnotesize
\centering
\caption{Results of the ART-based mixed-effects ANOVA for the post-trial questionnaire items.}
\label{tab:PTQ-ANOVA}
\renewcommand{\arraystretch}{1}
\begin{tabular}{crccrr@{\hspace{2pt}}l}
\toprule
DV & \multicolumn{1}{c}{Source} & \textit{df} & \textit{df.res} & \multicolumn{1}{c}{\textit{F}} & \multicolumn{2}{c}{\textit{p}} \\
\midrule
 & eHMI & 2 & 435.00 & 162.985 & $<$.001 & *** \\
Q1 & Takeover & 1 & 435.01 & 91.552 & $<$.001 & *** \\
 & Interaction & 2 & 435.02 & 1.346 & .261 &  \\
\midrule
 & eHMI & 2 & 435.03 & 78.958 & $<$.001 & *** \\
Q2 & Takeover & 1 & 435.03 & 33.500 & $<$.001 & *** \\
 & Interaction & 2 & 435.04 & 1.557 & .212 &  \\
\midrule
 & eHMI & 2 & 435.02 & 286.144 & $<$.001 & *** \\
Q3 & Takeover & 1 & 435.03 & 44.654 & $<$.001 & *** \\
 & Interaction & 2 & 435.03 & 1.563 & .211 &  \\
\midrule
 & eHMI & 2 & 435.02 & 1.852 & .158 &  \\
Q4 & Takeover & 1 & 435.04 & 592.281 & $<$.001 & *** \\
 & Interaction & 2 & 435.02 & 0.120 & .887 &  \\
\midrule
 & eHMI & 2 & 434.90 & 21.539 & $<$.001 & *** \\
Q5 & Takeover & 1 & 434.89 & 37.118 & $<$.001 & *** \\
 & Interaction & 2 & 434.90 & 3.336 & .037 & * \\
\midrule
 & eHMI & 2 & 434.98 & 8.282 & $<$.001 & *** \\
Q6 & Takeover & 1 & 434.98 & 52.465 & $<$.001 & *** \\
 & Interaction & 2 & 434.98 & 3.896 & .021 & * \\
\bottomrule
\multicolumn{7}{l}{\footnotesize DV: Dependent variable.  Takeover: Takeover outcomes of the AV.}\\
\multicolumn{7}{l}{\footnotesize *: $p<.05$, **: $p<.01$, ***: $p<.001$}
\end{tabular}
\end{table}

\begin{table}[!t]
\footnotesize
    \centering
\caption{FDR-corrected post hoc pairwise comparisons of eHMI conditions for the Q1, Q2 and Q3.}
    \label{tab:PTQ_eHMI_posthoc_Q1Q2Q3}
    \setlength\tabcolsep{6pt}
    \renewcommand{\arraystretch}{1}
\begin{tabular}{cccrrrr@{\hspace{2pt}}l}
\toprule
DV & eHMI(A) & eHMI(B) & \multicolumn{1}{c}{\textit{Est.}} & \multicolumn{1}{c}{\textit{SE}} & \textit{t}-ratio & \multicolumn{2}{c}{\textit{p}-corr} \\ \midrule
& C & N & 70.4 & 11.3 & 6.212 & $<$.001 & *** \\
Q1  & C+O & N & 201.7 & 11.3 & 17.788 & $<$.001 & *** \\ 
& C+O & C & 131.3 & 11.3 & 11.578 & $<$.001 & *** \\
\midrule
& C & N & 59.5 & 12.9 & 4.620 & $<$.001 & *** \\
Q2 & C+O & N & 160.2 & 12.9 & 12.431 & $<$.001 & *** \\ 
& C+O & C & 100.7 & 12.9 & 7.813 & $<$.001 & *** \\
\midrule
& C & N & 86.4 & 10.1 & 8.519 & $<$.001 & *** \\
Q3  & C+O & N & 239.7 & 10.2 & 23.620 & $<$.001 & *** \\ 
& C+O & C & 153.3 & 10.2 & 15.103 & $<$.001 & *** \\
   \bottomrule
\multicolumn{8}{l}{\footnotesize DV: Dependent variable. *: $p<.05$, **: $p<.01$, ***: $p<.001$}
\end{tabular}
\end{table}

As shown in Table~\ref{tab:PTQ_eHMI_posthoc_Q1Q2Q3}, all FDR-corrected pairwise comparisons among the three eHMI conditions were significant for Q1, Q2, and Q3 (all $p<.001$). 
For all three items, the ratings were highest under eHMI C+O, followed by eHMI C and then eHMI N, with all pairwise differences being significant ($p<.001$).
These results indicate that the proposed eHMI C+O most effectively improved participants' understanding of the AV's driving intention (Q1), prediction of the AV's behavior (Q2), and perceived sufficiency of the information presented by the AV (Q3).

\begin{table}[!t]
\footnotesize
\centering
\caption{Post hoc comparisons of takeover conditions for the Q1, Q2, Q3 and Q4.}
\label{tab:PTQ_takeover_posthoc_Q1Q2Q3Q4}
\setlength\tabcolsep{4pt}
\begin{tabular}{cccrrrr@{\hspace{2pt}}l}
\toprule
DV & Takeover(A) & Takeover(B) & \multicolumn{1}{c}{\textit{Est.}} & \multicolumn{1}{c}{\textit{SE}} & \textit{t}-ratio & \multicolumn{2}{c}{\textit{p}} \\
\midrule
Q1 & Successful & Unsuccessful & 104.0 & 10.8 & 9.568 & $<$.001 & *** \\
Q2 & Successful & Unsuccessful & 67.1 & 11.6 & 5.788 & $<$.001 & *** \\
Q3 &Successful & Unsuccessful & 76.1 & 11.4 & 6.682 & $<$.001 & *** \\
Q4 & Successful & Unsuccessful & -199.0 & 8.19 & -24.337 & $<$.001 & *** \\
\bottomrule
\multicolumn{8}{l}{\footnotesize DV: Dependent variable.  Takeover: Takeover outcomes of the AV.}\\
 \multicolumn{8}{l}{\footnotesize *: $p<.05$, **: $p<.01$, ***: $p<.001$.}
\end{tabular}
\end{table}

Post hoc comparisons in Table~\ref{tab:PTQ_takeover_posthoc_Q1Q2Q3Q4} show that ratings for understanding the AV’s driving intention (Q1), predicting the AV’s behavior (Q2), and perceived sufficiency of the information presented by the AV (Q3) were significantly higher in the successful takeover condition than in the unsuccessful takeover condition (all $p<.001$). 
Moreover, the rating for perceived danger from the AV’s behavior (Q4) was significantly lower in the successful takeover condition than in the unsuccessful takeover condition ($p<.001$).

\begin{table}[!t]
\footnotesize
\centering
\caption{FDR-corrected post hoc pairwise comparisons of eHMI conditions within each takeover condition for Q5 and Q6.}
\label{tab:PTQ_eHMI_posthoc_Q5Q6}
\setlength\tabcolsep{4pt}
\begin{tabular}{c@{\hspace{2pt}}c@{\hspace{2pt}}c@{\hspace{3pt}}c@{\hspace{2pt}}rrrr@{\hspace{2pt}}l}
\toprule
DV & Takeover & eHMI(A) & eHMI(B) & \multicolumn{1}{c}{\textit{Est.}} & \multicolumn{1}{c}{\textit{SE}} & \textit{t}-ratio & \multicolumn{2}{c}{\textit{p}-corr} \\
\midrule

\multirow{6}{*}{Q5}
& \multirow{3}{*}{Successful}
&  C & N & 2.85 & 7.53 & 0.379 & .705 & \\
& &C+O & C & 25.55 & 7.54 & 3.387 & .001 & ** \\
& & C+O & N & 28.40 & 7.54 & 3.765 & $<$.001 & *** \\

\cmidrule(l){2-9}
& \multirow{3}{*}{Unsuccessful}
&  C & N & 7.62 & 8.16 & 0.933 & .352 & \\
& &C+O & C & 31.17 & 8.16 & 3.819 & $<$.001 & *** \\
& & C+O & N & 38.79 & 8.16 & 4.752 & $<$.001 & *** \\

\midrule

\multirow{6}{*}{Q6}
& \multirow{3}{*}{Successful}
&  C & N & 14.5 & 9.37 & 1.544 & .186 & \\
& &C+O & C & -24.7 & 9.38 & -2.632 & .027 & * \\
& & C+O & N & -10.2 & 9.38 & -1.090 & .277 & \\

\cmidrule(l){2-9}
& \multirow{3}{*}{Unsuccessful}
&  C & N & -11.7 & 9.29 & -1.256 & .211 & \\
& &C+O & C & -23.5 & 9.29 & -2.528 & .018 & * \\
& & C+O & N & -35.1 & 9.29 & -3.784 & .001 & *** \\
\bottomrule
\multicolumn{8}{l}{\footnotesize DV: Dependent variable.  Takeover: Takeover outcomes of the AV.}\\
\multicolumn{9}{l}{\footnotesize *: $p<.05$, **: $p<.01$, ***: $p<.001$.}
\end{tabular}
\end{table}

Because significant interaction effects between eHMI and takeover were found for Q5 and Q6 in the ART ANOVA (see Table~\ref{tab:PTQ-ANOVA}), post hoc comparisons were conducted separately within each takeover condition (see Table~\ref{tab:PTQ_eHMI_posthoc_Q5Q6}).
For Q5, ratings for confidence in handling the driving operation were significantly higher under eHMI C+O than under both eHMI C and eHMI N in both the successful and unsuccessful takeover conditions ($p < .001$), while no significant difference was found between eHMI C and eHMI N in either takeover condition ($p= .352$).
For Q6, the lowest ratings were observed under eHMI C+O, indicating that participants perceived the least hesitation in responding to the preceding AV’s RtI under this condition. 
This pattern was significant relative to eHMI C($p=.027$) in the successful takeover condition, and relative to both eHMI C ($p=.018$) and eHMI N ($p=.001$) in the unsuccessful takeover condition.

\subsection{Overall Evaluation via Post-experiment Questions}

\begin{figure}[!t]
    \centering
    \centerline{
        \subfloat[The plot of the E1 result.]{\includegraphics[width=0.5\linewidth]{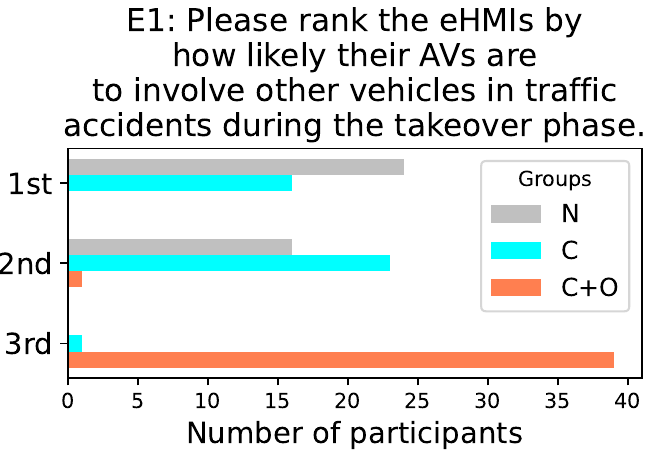}
        \label{fig:E1}}
        \hfill
        \subfloat[The plot of the E2 result.]{\includegraphics[width=0.5\linewidth]{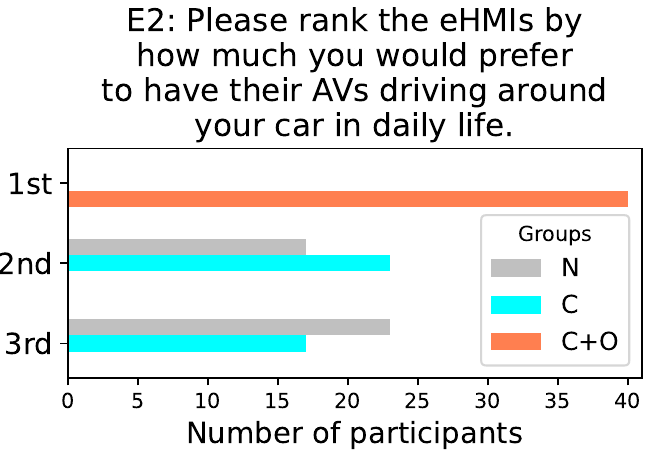}
        \label{fig:E2}}
        }
    \caption{The results of the post-experiment questionnaire.}
    \label{fig:PEQ}

\vspace{4mm}
   
    \centering
     \footnotesize
     \setlength\tabcolsep{6pt}
    \captionof{table}{Post hoc pairwise comparisons by Wilcoxon signed-rank test (two-sided) with FDR correction.}
    \label{tab:E-post}
    \renewcommand{\arraystretch}{1}
\begin{tabular}{ccccr@{\hspace{1pt}}lr}
\toprule
Questions & eHMI(A) & eHMI(B) & \textit{W}-val & \multicolumn{2}{c}{\textit{p}-corr} & \textit{RBC} \\ 
\midrule
\multirow{3}{*}{E1}
& C   & N   & 320.0 & .172    &     & -0.220 \\
& C+O & N   & 0.0   & $<$.001 & *** & -1.000 \\
& C+O   & C & 12.5  & $<$.001 & *** & -0.970 \\
\midrule
\multirow{3}{*}{E2}
& C   & N   & 348.5 & .347    &     & 0.150 \\
& C+O & N   & 0.0   & $<$.001 & *** & 1.000 \\
& C+O   & C & 0.0   & $<$.001 & *** & 1.000 \\
\bottomrule
\multicolumn{7}{l}{\footnotesize  Rank coding: 1st place = 3, 2nd place = 2, 3rd place = 1. }\\
\multicolumn{7}{l}{\footnotesize  \textit{RBC}: Matched pairs rank-biserial correlation. ***: $p<.001$}
\end{tabular}
\end{figure}

Figure~\ref{fig:PEQ} presents the overall rankings of the three eHMI conditions for two post-experiment questions, \ie E1 and E2 (see section~\ref{sec:PEQ}). 
Participants ranked the eHMI conditions from 1st place (most likely) to 3rd place (least likely).

Friedman tests revealed significant differences among the three eHMI conditions for both E1 ($\chi^2(2) = 58.05$, $p < .001$, Kendall's $W = 0.726$), and E2 ($\chi^2(2) = 60.45$, $p < .001$, Kendall's $W = 0.756$).
Furthermore, the results of post hoc pairwise comparisons using Wilcoxon signed-rank tests with FDR correction are shown in Table~\ref{tab:E-post}.
The results showed that, in E1, AVs equipped with eHMI N or eHMI C were ranked as significantly more likely to involve other vehicles in traffic accidents during the takeover phase than AVs equipped with eHMI C+O (both $p<.001$), while no significant difference was found between AVs equipped with eHMI N and those equipped with eHMI C ($p=.172$).
In E2, participants significantly preferred AVs equipped with eHMI C+O to be driving around their cars in daily life compared with AVs equipped with eHMI N or eHMI C (both $p<.001$), while no significant difference was found between AVs equipped with eHMI N and those equipped with eHMI C.

\subsection{Time Headway Changes After the AV's RtI Issued}

\begin{figure}[!p]
    \centering
    \begin{subfigure}[b]{0.49\linewidth}
        \centering
        \includegraphics[width=\linewidth,trim=5 5 5 5, clip]{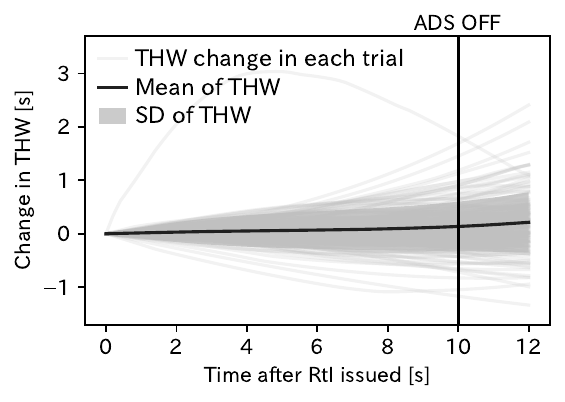}
        \caption{THW Changes under eHMI N}
        \label{fig:THW_diff_N_All}
    \end{subfigure}
    \hfill
    \begin{subfigure}[b]{0.49\linewidth}
        \centering
        \includegraphics[width=\linewidth,trim=5 5 5 5, clip]{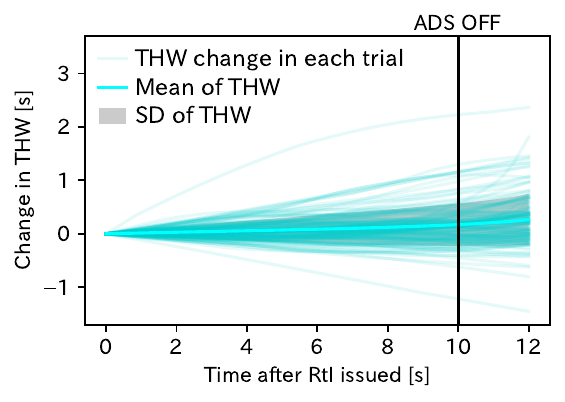}
        \caption{THW changes under eHMI C}
        \label{fig:THW_diff_C_All}
    \end{subfigure}
     \\
     \vspace{2mm}
\begin{subfigure}[b]{0.49\linewidth}
        \centering
        \includegraphics[width=\linewidth,trim=5 5 5 5, clip]{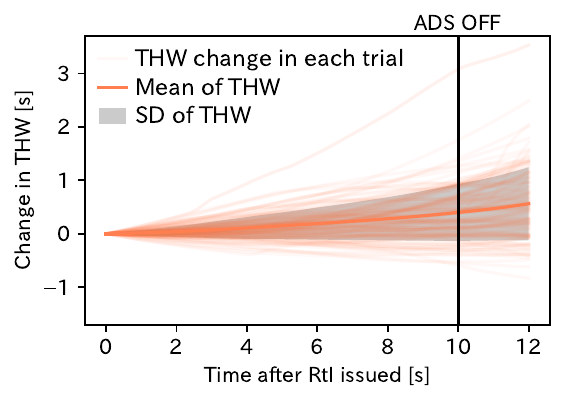}
        \caption{THW changes under eHMI C+O}
        \label{fig:THW_diff_C+O_All}
    \end{subfigure}
    \hfill
    \begin{subfigure}[b]{0.49\linewidth}
        \centering
        \includegraphics[width=\linewidth,trim=5 5 5 0, clip]{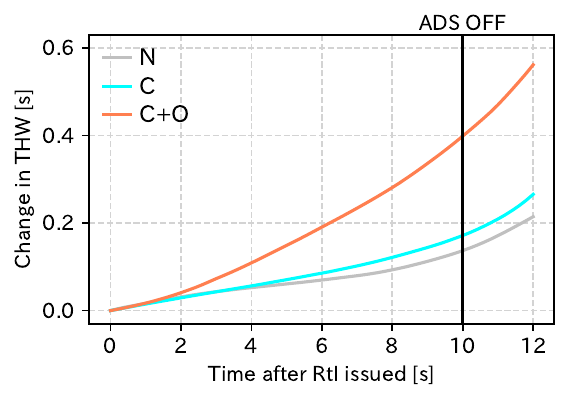}
        \caption{Mean of THW changes for eHMI conditions}
        \label{fig:THW_diff_all_All}
    \end{subfigure}
  
    \caption{Changes in THW after the AV issued the RtI under different eHMI conditions.}
    \label{fig:THW}

\vspace{2mm}

\footnotesize
\centering
\captionof{table}{ART mixed-effects ANOVA results for the AUC of THW change after the AV's RtI.}
\label{tab:auc_anova}
\setlength\tabcolsep{7pt}
\begin{tabular}{crccrr@{\hspace{2pt}}l}
\toprule
DV & Source & \textit{df} & \textit{df.res} & \multicolumn{1}{c}{\textit{F}} & \multicolumn{2}{c}{\textit{p}} \\
\midrule
$\texttt{AUC}^{\texttt{THW}}_{1s}$
& eHMI        & 2 & 438 & 0.603  & .547 &    \\
\midrule
$\texttt{AUC}^{\texttt{THW}}_{2s}$
& eHMI        & 2 & 438 & 1.229  & .294 &    \\
\midrule
$\texttt{AUC}^{\texttt{THW}}_{3s}$
& eHMI        & 2 & 438 & 2.159  & .117 &    \\
\midrule
$\texttt{AUC}^{\texttt{THW}}_{4s}$
& eHMI        & 2 & 438 & 3.375  & .035 & *  \\
\midrule
$\texttt{AUC}^{\texttt{THW}}_{5s}$
& eHMI        & 2 & 438 & 4.930  & .008 & ** \\
\midrule
$\texttt{AUC}^{\texttt{THW}}_{10s}$
& eHMI        & 2 & 438 & 12.202 & $<$.001 & *** \\
\midrule
\multirow{3}{*}{$\texttt{AUC}^{\texttt{THW}}_{12s}$}
& eHMI        & 2 & 435 & 15.195 & $<$.001 & *** \\
& Takeover    & 1 & 435 & 2.958  & .086 &    \\
& Interaction & 2 & 435 & 0.379  & .685 &    \\
\bottomrule
\multicolumn{7}{l}{\footnotesize DV: Dependent variable.  Takeover: Takeover outcomes of the AV.}\\
\multicolumn{7}{l}{\footnotesize*: $p<.05$, **: $p<.01$, ***: $p<.001$}
\end{tabular}

\vspace{4mm}

\footnotesize
\centering
\captionof{table}{FDR-corrected post hoc pairwise comparisons for $\texttt{AUC}^{\texttt{THW}}_{T\in\{1,2,3,4,5,10\}s}$ under the three eHMI conditions based on the ART mixed-effects models.}
\label{tab:auc_posthoc}
\setlength\tabcolsep{4pt}
\begin{tabular}{cccrrrr@{\hspace{2pt}}l}
\toprule
DV & eHMI (A) & eHMI (B) & \textit{Est.} & \textit{SE} & \multicolumn{1}{c}{$t$} & \multicolumn{2}{c}{$p$-corr} \\
\midrule
\multirow{3}{*}{$\texttt{AUC}^{\texttt{THW}}_{4s}$}
& C   & N   &  5.95 & 14.7 & 0.406 & .685 & \\
& C+O & N   & 35.60 & 14.7 & 2.425 & .047 & * \\
& C+O & C   & 29.64 & 14.7 & 2.020 & .066 & \\
\midrule
\multirow{3}{*}{$\texttt{AUC}^{\texttt{THW}}_{5s}$}
& C   & N   &  6.47 & 14.6 & 0.443 & .658 & \\
& C+O & N   & 42.53 & 14.6 & 2.914 & .011 & * \\
& C+O & C   & 36.06 & 14.6 & 2.471 & .021 & * \\
\midrule
\multirow{3}{*}{$\texttt{AUC}^{\texttt{THW}}_{10s}$}
& C   & N   &  7.23 & 14.2 & 0.508 & .612 & \\
& C+O & N   & 64.20 & 14.2 & 4.510 & $<$.001 & *** \\
& C+O & C   & 56.97 & 14.2 & 4.002 & $<$.001 & *** \\
\midrule
\multirow{3}{*}{$\texttt{AUC}^{\texttt{THW}}_{12s}$}
& C   & N   &  7.22 & 14.1 & 0.513 & .608 & \\
& C+O & N   & 70.51 & 14.1 & 5.010 & $<$.001 & *** \\
& C+O & C   & 63.29 & 14.1 & 4.497 & $<$.001 & *** \\
\bottomrule
\end{tabular}
\end{figure}

Fig.~\ref{fig:THW} presents the time-series changes in THW after the AV issued the RtI under the three eHMI conditions. 
Figs.~\ref{fig:THW_diff_N_All}, \ref{fig:THW_diff_C_All}, and \ref{fig:THW_diff_C+O_All} present the THW change trajectories for each trial together with the mean and standard deviation under eHMI N, eHMI C, and eHMI C+O, respectively, while Fig.~\ref{fig:THW_diff_all_All} compares the mean THW changes across the three eHMI conditions.
As shown in Fig.~\ref{fig:THW_diff_all_All}, the mean THW increased over time under all three eHMI conditions after the RtI was issued.
During the early phase, the three conditions showed similar trends, and the differences among them were small. 
Around 2~s after the RtI, the mean THW under the eHMI C+O condition began to increase more markedly and gradually exceeded those under the other two conditions.
The gap among the conditions continued to widen over time, with the eHMI C+O condition showing the largest increase in mean THW even before ADS deactivation at 10~s, and this difference became more pronounced afterward.
In contrast, the mean THW under the eHMI N and C conditions increased more gradually and remained relatively similar, although the eHMI C condition was slightly higher than the eHMI N condition in the later phase.

To further quantify this pattern, the AUC of THW change in the MV from the onset of the AV's RtI to each time point was analyzed (see Table~\ref{tab:auc_anova}).
No significant effect of eHMI was found for $\texttt{AUC}^{\texttt{THW}}_{1s}$, $\texttt{AUC}^{\texttt{THW}}_{2s}$ and $\texttt{AUC}^{\texttt{THW}}_{3s}$. 
In contrast, the main effect of eHMI became significant from 4~s after the RtI onward, including $\texttt{AUC}^{\texttt{THW}}_{4s}$ ($F(2,438)=3.375$, $p=.035$), $\texttt{AUC}^{\texttt{THW}}_{5s}$ ($F(2,438)=4.930$, $p=.008$), $\texttt{AUC}^{\texttt{THW}}_{10s}$ ($F(2,438)=12.202$, $p<.001$), and $\texttt{AUC}^{\texttt{THW}}_{12s}$ ($F(2,435)=15.195$, $p<.001$). 
This result suggests that the effect of eHMI on participants' THW adjustment emerged after several seconds and became increasingly pronounced over time. 
For $\texttt{AUC}^{\texttt{THW}}_{12s}$, no significant main effect of the takeover outcome of the AV or interaction effect was found.

Post hoc pairwise comparisons further showed that the significant eHMI effect from 4~s onward was mainly driven by the eHMI C+O condition (see Table~\ref{tab:auc_posthoc}). 
For $\texttt{AUC}^{\texttt{THW}}_{4s}$, eHMI C+O was significantly higher than eHMI N ($p=.047$), while the differences between eHMI C and N and between eHMI C+O and C were not significant. 
For $\texttt{AUC}^{\texttt{THW}}_{5s}$, eHMI C+O was significantly higher than both eHMI N ($p=.011$) and eHMI C ($p=.021$), while no significant difference was found between eHMI C and N. 
The same pattern was observed for $\texttt{AUC}^{\texttt{THW}}_{10s}$ and $\texttt{AUC}^{\texttt{THW}}_{12s}$, where eHMI C+O was significantly higher than both eHMI N and eHMI C (all $p<.001$), but no significant difference was found between eHMI C and N. 
These results indicate that the increase in THW change was specifically enhanced under the proposed eHMI C+O condition, and that this advantage became clearer as time elapsed after the RtI.

\subsection{MV Accident Occurrence in the AV Accident Scenario}

Table~\ref{tab:mv_accident_counts} presents the numbers of trials with and without MV accidents under each eHMI condition in the AV accident scenario. 
The observed accident rates were 20.0\% under eHMI N, 13.8\% under eHMI C, and 6.3\% under eHMI C+O. 
Thus, descriptively, the accident rate decreased in the order of eHMI N, eHMI C, and eHMI C+O. 
Fisher's exact test showed a significant difference in MV accident occurrence among the three eHMI conditions ($p=.032$).

\begin{table}[!t]
\footnotesize
\centering
\caption{Numbers of trials with and without MV accidents under each eHMI condition in the AV accident scenario.}
\label{tab:mv_accident_counts}
\begin{tabular}{ccccr}
\toprule
 eHMI & \begin{tabular}[c]{@{}c@{}}Accident\\ trials\end{tabular} & \begin{tabular}[c]{@{}c@{}}Non-accident\\ trials\end{tabular} & \begin{tabular}[c]{@{}c@{}}Total\\ trials\end{tabular} & \multicolumn{1}{c}{\begin{tabular}[c]{@{}c@{}}Accident\\ rate\end{tabular}} \\ 
 \midrule
 N & 16 & 64 & 80 & 20.0\% \\
  C & 11 & 69 & 80 & 13.8\% \\
  C+O & 5 & 75 & 80 & 6.3\% \\ 
 \bottomrule
\end{tabular}

\vspace{4mm}

\footnotesize
\centering
\caption{Fitting result of the binomial generalized linear mixed model for MV accident occurrence.}
\label{tab:glmm_mv_accident}
\setlength\tabcolsep{4pt}
\begin{tabular}{clrcrrr@{\hspace{2pt}}l}
\toprule
Model & Source & \multicolumn{1}{c}{\textit{Est.}} & \multicolumn{1}{c}{\textit{OR}} & \multicolumn{1}{c}{\textit{SE}} & \multicolumn{1}{c}{$z$} & \multicolumn{2}{c}{$p$} \\
\midrule
\multirow{3}{*}{Model 1} 
& Intercept & -1.689 & 0.185 & 0.393 & -4.299 & $<$.001 & *** \\
& eHMI C   & -0.513 & 0.599 & 0.459 & -1.118 & .263 & \\
& eHMI C+O & -1.461 & 0.232 & 0.571 & -2.561 & .010 & * \\
\bottomrule
\multicolumn{7}{l}{\footnotesize Participant ID was included as a random intercept.}\\
\multicolumn{7}{l}{\footnotesize The reference level of eHMI condition was the eHMI N.}\\
\multicolumn{7}{l}{\footnotesize \textit{Est.}: Estimate on the log-odds scale. \textit{OR}: odds ratio = $\exp(\textit{Est.})$.}\\
\multicolumn{7}{l}{\footnotesize  **: $p < .01$, ***: $p < .001$.}
\end{tabular}

\vspace{4mm}

\footnotesize
\centering
\setlength\tabcolsep{6pt}
\caption{FDR-corrected post hoc pairwise comparisons among eHMI conditions for MV accident occurrence based on the binomial generalized linear mixed model.}
\label{tab:mv_accident_posthoc}
\begin{tabular}{ccrrcrr@{\hspace{2pt}}l}
\toprule
eHMI (A) & eHMI (B) & \textit{Est.} & \textit{OR} & \textit{SE} & \multicolumn{1}{c}{\textit{z}} & \multicolumn{2}{c}{$p$-corr} \\
\midrule
C   & N   & -0.513 & 0.599 & 0.459 & -1.118 & .263 &   \\
C+O & N   & -1.461 & 0.232 & 0.571 & -2.561 & .031 & * \\
C+O & C   & -0.948 & 0.387 & 0.589 & -1.610 & .161 &   \\
\bottomrule
\multicolumn{8}{l}{\footnotesize \textit{Est.}: Estimate on the log-odds scale. \textit{OR}: odds ratio = $\exp(\textit{Est.})$.}\\
\multicolumn{8}{l}{\footnotesize \textit{SE}: Standard error. * $p<.05$.}
\end{tabular}
\vspace{-2mm}
\end{table}

To further examine the overall effect of eHMI on MV accident occurrence, Model 1 in Table~\ref{tab:glmm_mv_accident}, which served as the primary model (Eq.~\ref{eq:main_model}), was specified as a binomial generalized linear mixed model with eHMI as a fixed effect and participant ID as a random intercept.
Using eHMI N as the reference level, the coefficient for eHMI C was not significant ($\mathrm{OR}=0.599$, $p=.263$), while the coefficient for eHMI C+O was significant ($\mathrm{OR}=0.232$, $p=.010$). 
Because an odds ratio (OR) smaller than 1 indicates lower odds of MV accident occurrence relative to the reference condition, the OR of 0.232 means that the odds under eHMI C+O were 23.2\% of those under eHMI N, corresponding to a 76.8\% reduction in accident odds.

FDR-corrected post hoc pairwise comparisons further confirmed that the odds of MV accident occurrence were significantly lower under eHMI C+O than under eHMI N ($\mathrm{OR}=0.232$, $p=.031$; Table~\ref{tab:mv_accident_posthoc}). 
In contrast, no significant difference was found between eHMI C and eHMI N or between eHMI C+O and eHMI C. 
These results indicate that the proposed eHMI C+O was associated with the lowest likelihood of MV accident occurrence among the three eHMI conditions.

\subsection{Exploratory Underlying Mechanisms Analysis of the eHMI Effect on MV Accident}

\begin{figure}[!t]
    \centering
    \includegraphics[width=1\linewidth]{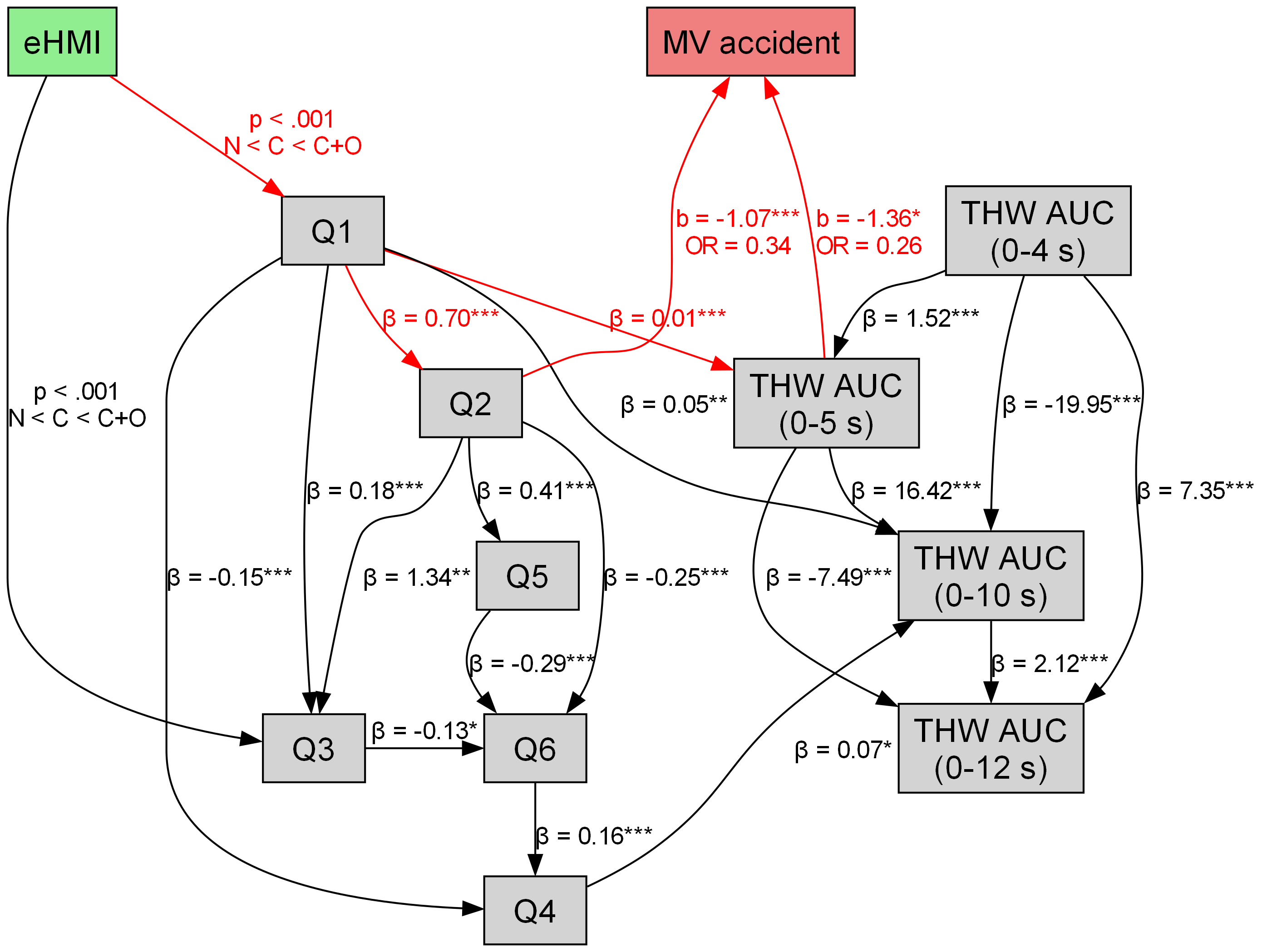}
    \caption{Final exploratory piecewise structural equation model showing the candidate dependency structure from eHMI condition to MV accident occurrence through subjective evaluations and THW AUC measures. Red arrows indicate the main paths associated with eHMI conditions and MV accident occurrence.}
    \label{fig:psem}
    \vspace{-4mm}
\end{figure}

\begin{table*}[!t]
\footnotesize
\centering
\caption{Component models included in the exploratory piecewise structural equation model.}
\label{tab:piecewise_component_models}
\renewcommand{\arraystretch}{1.5}
\begin{tabular}{llll}
\toprule
 & Component model & Type & Model formula \\
\midrule
\multirow{4}{*}{\shortstack{Main\\component\\models}}
& $\texttt{M}_\texttt{Q1}$ & LMM &
\texttt{Q1 $\sim$ eHMI + (1|ID)} \\
& $\texttt{M}_\texttt{Q2}$ & LMM &
\texttt{Q2 $\sim$ eHMI + Q1 + (1|ID)} \\
& $\texttt{M}_\texttt{AUC5s}$ & LMM &
\texttt{$\texttt{AUC}^{\texttt{THW}}_{5s}$ $\sim$ eHMI + Q1 + Q2 + (1|ID)} \\
& $\texttt{M}_\texttt{ACC}$ & Bi-GLMM &
\texttt{MV$_{Accident}$ $\sim$ eHMI + Q1 + Q2 + $\texttt{AUC}^{\texttt{THW}}_{5s}$ + (1|ID)} \\
\midrule
\multirow{7}{*}{\shortstack{Additional\\component\\models}}
& $\texttt{M}_\texttt{Q3}$ & LMM &
\texttt{Q3 $\sim$ eHMI + Q2 + Q1 + (1|ID)} \\
& $\texttt{M}_\texttt{Q4}$ & LMM &
\texttt{Q4 $\sim$ Q1 + Q6 + (1|ID)} \\
& $\texttt{M}_\texttt{Q5}$ & LMM &
\texttt{Q5 $\sim$ Q2 + (1|ID)} \\
& $\texttt{M}_\texttt{Q6}$ & LMM &
\texttt{Q6 $\sim$ Q2 + Q3 + Q5 + (1|ID)} \\
& $\texttt{M}_\texttt{AUC4s}$ & LMM &
\texttt{$\texttt{AUC}^{\texttt{THW}}_{4s}$ $\sim$ 1 + (1|ID)} \\
& $\texttt{M}_\texttt{AUC10s}$ & LMM &
\texttt{$\texttt{AUC}^{\texttt{THW}}_{10s}$ $\sim$ $\texttt{AUC}^{\texttt{THW}}_{4s}$ + $\texttt{AUC}^{\texttt{THW}}_{5s}$ + Q1 + Q4 + (1|ID)} \\
& $\texttt{M}_\texttt{AUC12s}$ & LMM &
\texttt{$\texttt{AUC}^{\texttt{THW}}_{12s}$ $\sim$ $\texttt{AUC}^{\texttt{THW}}_{4s}$ + $\texttt{AUC}^{\texttt{THW}}_{5s}$ + $\texttt{AUC}^{\texttt{THW}}_{10s}$ + (1|ID) }\\
\bottomrule
\addlinespace[-2pt]
\multicolumn{4}{l}{\footnotesize LMM: linear mixed-effects model. Bi-GLMM: binomial generalized linear mixed-effects model.} 
\end{tabular}
\end{table*}

The path structure of the pSEM was specified based on the candidate dependency structure obtained from the exploratory Gaussian Bayesian network analysis.
The final exploratory pSEM (\ie overall model), shown in Fig.~\ref{fig:psem}, represents an approximate dependency structure linking the eHMI condition to MV accident occurrence through subjective evaluations and THW-based behavioral response variables. 
The overall model fit was acceptable, with Fisher’s $C = 90.356$, $df = 86$, and $p = .353$.

The specific component models included in this pSEM are summarized in Table~\ref{tab:piecewise_component_models}.
Among these component models, four were defined as the main component models,\ie component models of $\texttt{M}_\texttt{Q1}$, $\texttt{M}_\texttt{Q2}$, $\texttt{M}_\texttt{AUC5s}$ and $\texttt{M}_\texttt{ACC}$, because they represented the primary exploratory pathway linking the eHMI condition to MV accident occurrence through subjective understanding and THW-based behavioral responses, as indicated by the red paths in Fig.~\ref{fig:psem}.
The remaining models were included as additional component models to represent the broader dependency structure suggested by the exploratory Bayesian network analysis, as indicated by the black paths in Fig.~\ref{fig:psem}.
In Fig.~\ref{fig:psem}, the value shown next to each path is the unstandardized estimated coefficient ($\beta$) because the original data structure was retained and the variables were not standardized.
For the component model of MV accident, both the log-odds coefficients ($b$) and the corresponding OR are presented.
Variance inflation factor (VIF) diagnostics showed no problematic multicollinearity in the main component models and in most of the additional component model, with all corresponding VIF values below the threshold of 4.0~\citep{o2007caution}..
Although high multicollinearity was observed among the THW AUC variables in the additional component models of $\texttt{M}_\texttt{AUC10s}$ and $\texttt{M}_\texttt{AUC12s}$, this was expected because they were derived from overlapping time windows and did not affect the main interpretation of the model.

\section{DISCUSSION}

\subsection{Effects of the proposed eHMI on surrounding drivers' understanding and subjective feelings}

One important finding is that the proposed eHMI C+O consistently improved participants' subjective understanding of the preceding AV. 
As shown in Table~\ref{tab:PTQ_eHMI_posthoc_Q1Q2Q3}, for Q1, Q2, and Q3, all pairwise differences among the three eHMI conditions were significant, with the ratings following the order of eHMI C+O, eHMI C, and eHMI N. 

These results indicate that the proposed eHMI C+O most effectively helped surrounding drivers understand the AV's driving intention, predict its subsequent behavior, and perceive that the information presented by the AV was sufficient.
These results indicate that, after the AV issued an RtI, externally presenting ADS status information through an eHMI, as in eHMI C and eHMI C+O, helped surrounding MV drivers better understand the AV's driving intention and predict its subsequent behavior compared with the no-eHMI condition (eHMI N).
This result is not unexpected, because similar findings and conclusions have been reported in many previous studies examining AV-MV interaction~\citep{li2023av,hub2023promoting} and AV-pedestrian interaction through eHMI~\citep{liu2025pre,liu2025data, kato2026investigating}.

More importantly, compared with eHMI C, which corresponds to the \textit{Automated Driving Marker Lights} currently being introduced to indicate ADS activation status, the proposed eHMI C+O yielded significantly better ratings on Q1, Q2, and Q3.
A plausible explanation is that the proposed eHMI C+O provided information that was more directly relevant to the surrounding driver's immediate decision-making during the takeover phase. 
For a surrounding MV driver, it may not be sufficient to know only that the lead vehicle is currently under ADS control. 
More critical is whether the AV has entered a safety-critical transition phase after issuing an RtI and whether the driver has already resumed control.

The results for Q5 and Q6 (see Table~\ref{tab:PTQ_eHMI_posthoc_Q5Q6}) further suggest that the proposed eHMI C+O supported not only cognitive understanding but also participants' subjective readiness to respond to the AV.
For Q5, confidence in handling the driving operation was significantly higher under eHMI C+O than under both eHMI C and eHMI N in both takeover conditions.
For Q6, the lowest hesitation ratings were observed under eHMI C+O, with significant advantages over eHMI C in both takeover conditions and over eHMI N in the unsuccessful takeover condition.
These results suggest that the proposed eHMI C+O did not merely make MV drivers feel more informed, but also helped them respond with less hesitation and greater confidence during the AV's takeover phase.
Notably, the advantage in Q6 was more pronounced in the AV unsuccessful takeover condition, where the traffic risk was higher.
This finding suggests that explicitly presenting RtI issuance and driver takeover status may be particularly valuable when the surrounding driver faces greater uncertainty and urgency.

\subsection{Effects of the proposed eHMI on early defensive driving behavior}

Beyond subjective evaluations, the THW results indicate that the proposed eHMI C+O also affected surrounding drivers' actual driving behavior after the AV issued the RtI.
The ART ANOVA analysis of the THW AUC measures supports this interpretation, as shown in Table~\ref{tab:auc_anova}.
No significant eHMI effect was found for $\texttt{AUC}^{\texttt{THW}}_{1s}$, $\texttt{AUC}^{\texttt{THW}}_{2s}$, or $\texttt{AUC}^{\texttt{THW}}_{3s}$, but significant main effects of eHMI emerged from 4~s onward.
Post hoc comparisons (see Table~\ref{tab:auc_posthoc}) showed that this effect was mainly driven by the proposed eHMI C+O, which yielded significantly larger THW increases than eHMI N from 4~s onward, and significantly larger increases than both eHMI N and eHMI C from 5~s onward. 
In contrast, no significant difference was found between eHMI C and eHMI N. These results suggest that the proposed eHMI C+O promoted an earlier defensive increase in following distance, whereas merely indicating ADS activation was not sufficient to reliably induce such behavioral adaptation.

This temporal pattern is also meaningful. 
The absence of a significant eHMI effect during the first 1--3~s may reflect the time required for surrounding drivers to perceive the signal, interpret its meaning, and initiate a behavioral response. 
Once this processing period had passed, the behavioral effect of the proposed eHMI became evident and then accumulated over time. 
Although we are not aware of any study that has directly quantified surrounding MV drivers' response time to warning signals issued by a nearby AV, the observed 1--3~s interval may be broadly consistent with the takeover reaction time reported for Level 3 AV drivers without an NDRT~\citep{gold2016taking,yoon2019effects}.

Another noteworthy finding is that no significant difference in THW change was found between the eHMI C and eHMI N conditions over the entire 0--12~s period (see Tables~\ref{tab:auc_anova} and~\ref{tab:auc_posthoc}). 
This result is consistent with \citet{stange2022manual}, who likewise reported that the presence or absence of an eHMI similar to the \textit{Automated Driving Marker Lights} did not significantly influence manually driven vehicles' THW in mixed traffic with different AV penetration rates. 

In summary, these findings suggest that indicating ADS activation status alone, as in the \textit{Automated Driving Marker Lights}, may be insufficient to reliably influence surrounding MV drivers' defensive driving behavior during the AV's RtI phase. 
By contrast, the proposed eHMI C+O produced a clear behavioral advantage, further supporting the importance of additionally presenting the RtI state and the driver's takeover state.

\subsection{Association with reduced MV accident occurrence}

The accident analysis further suggests that the benefit of the proposed eHMI C+O extended beyond subjective evaluations and THW adjustment to a safety-relevant outcome. 
Descriptively, the MV accident rate decreased from 20.0\% under eHMI N to 13.8\% under eHMI C and to 6.3\% under eHMI C+O (see Table~\ref{tab:mv_accident_counts}). 
The post hoc comparisons in Table~\ref{tab:mv_accident_posthoc} further showed that, relative to eHMI N, the odds of MV accident occurrence were significantly lower under eHMI C+O ($\mathrm{OR}=0.232$, $p=.031$).
This means that the accident odds under eHMI C+O were significantly reduced by 76.8\% relative to eHMI N. 
Although the difference between eHMI C+O and eHMI C did not reach statistical significance, the odds ratio for this comparison was 0.387, indicating that the accident odds under eHMI C+O were 38.7\% of those under eHMI C, corresponding to a numerical 61.3\% reduction in accident odds.

Importantly, the effect was significant for eHMI C+O but not for eHMI C.
This pattern is consistent with the questionnaire and THW results and strengthens the interpretation that the safety benefit was not produced merely by indicating that the lead vehicle was automated. Rather, the critical factor appears to be the explicit communication of the RtI-related takeover status. By helping surrounding drivers better understand the AV's situation and begin defensive responses earlier, the proposed eHMI C+O may have reduced the likelihood that the surrounding MV would be involved in a collision when the AV driver failed to take over successfully.

\subsection{Exploratory analysis of underlying mechanisms of the eHMI
effect on MV accident}

Because MV involvement occurred in only a small proportion of the AV accident cases (see Table~\ref{tab:mv_accident_counts}), the available data were not well suited for a rigorous causal analysis. Accordingly, this part was treated as an exploratory analysis, and the results were discussed only as a supplement to the main findings.

In the final exploratory pSEM (see Fig.~\ref{fig:psem}), eHMI condition was not directly linked to MV accident occurrence, but was instead associated with accident risk mainly through a subjective pathway and an early behavioral-response pathway. 
Specifically, eHMI condition showed a significant overall effect on Q1, with the estimated values ordered as C+O, C, and N. 
This result suggests that the proposed eHMI C+O enabled surrounding drivers to better understand the AV’s current state and driving intention during the AV RtI period.
In turn, Q1 was positively associated with Q2, suggesting that improved understanding of the AV’s driving intention helped surrounding MV drivers better predict its subsequent driving behavior.
This is consistent with the situation awareness theory~\citep{endsley1995toward}, particularly the link between comprehension of the current situation and projection of future states, and is also in line with previous findings in interactions between pedestrians and eHMI-equipped AVs~\citep{liu2025pre,liu2025data}.
Meanwhile, Q1 was positively associated with $\texttt{AUC}^{\texttt{THW}}_{5s}$ rather than $\texttt{AUC}^{\texttt{THW}}_{10s}$ or $\texttt{AUC}^{\texttt{THW}}_{12s}$. 
This suggests that improved understanding of the AV’s driving intention enabled surrounding MV drivers to increase their following distance at an earlier stage after the RtI was issued, thereby promoting a more defensive driving response.
In the final component model of MV accident, only Q2 and $\texttt{AUC}^{\texttt{THW}}_{5s}$ showed significant direct effects on MV accident occurrence, and both effects were negative (Q2: $\beta=-1.07$, OR $=0.34$; $\texttt{AUC}^{\texttt{THW}}_{5s}$: $\beta=-1.37$, OR $=0.26$). 
This also suggests that, once the AV issues an RtI, preventing the surrounding MV from becoming involved in the AV’s accident may depend on whether the MV driver can both better predict the AV’s subsequent behavior and increase the following distance at an earlier stage.

In summary, these findings suggest that the proposed eHMI may be associated with lower accident risk not through a direct effect, but rather by improving surrounding MV drivers’ situation awareness of the AV and by promoting an earlier defensive driving response.
As this analysis was exploratory, these paths should be interpreted as approximate dependency paths rather than as confirmed causal mediation effects.

\subsection{Limitations and future work}

Several limitations should be noted. 
First, this study was conducted in a driving simulator, and the extent to which the observed effects generalize to real-world traffic environments remains to be confirmed.
Real traffic includes richer visual context, greater variability in vehicle behavior, and potentially different levels of time pressure and distraction. 

Second, the participants received pre-instruction about the meaning of each eHMI condition before the experiment.
Although this pre-instruction helped ensure that participants understood the information conveyed by the eHMI during the experiment, surrounding drivers in real traffic may not always have such prior knowledge.
Future studies should therefore examine how intuitively the proposed eHMI can be interpreted without prior explanation or prior knowledge and whether learning effects occur over repeated exposure.

Third, MV accident occurrence was a relatively rare event in the present dataset, resulting in sparse accident data. 
Such data sparsity can affect the stability of logistic-regression estimates, especially when the analysis is extended from overall effects to potential explanatory pathways. 
For this reason, the analysis of the potential mechanism linking eHMI and MV accident occurrence was treated as exploratory rather than confirmatory. 
Accordingly, the identified paths should be interpreted as tentative dependency patterns, and future studies with larger samples and more accident cases are needed to verify the robustness of these relationships.

Fourth, the present study mainly focused on planned RtI situations, in which the AV issued the RtI with a relatively sufficient lead time before reaching its system limit. 
In more sudden RtI situations, the time available for the AV driver to respond may be shorter, which would also reduce the time during which the eHMI can warn surrounding drivers. 
As a result, the effectiveness of the proposed eHMI may differ under more time-critical RtI conditions. 
Future studies should therefore examine whether the proposed eHMI remains effective when the available warning time is reduced, and identify the minimum lead time required for it to provide meaningful support to surrounding drivers.

Despite these limitations, the present findings provide evidence that presenting RtI-related takeover status through an eHMI may help surrounding drivers better understand Level~3 AVs, respond more defensively during the takeover phase, and reduce the risk of accident involvement in MV--AV mixed traffic.

\section{CONCLUSION}

This study examined whether an eHMI that explicitly presents the RtI-related status of a Level~3 AV to surrounding vehicles can help surrounding MV drivers better understand the AV and respond more safely during the takeover phase.
Overall, the results suggest that the proposed eHMI C+O presenting the RtI-related takeover status was the most effective among the three eHMI conditions. 
Compared with eHMI N and eHMI C, it improved participants' understanding of the AV's driving intention, prediction of the AV's behavior, and perceived sufficiency of the information presented by the AV. 
It also reduced hesitation in responding to the preceding AV's RtI and increased confidence in handling the driving operation. 
Importantly, these subjective benefits were accompanied by earlier and larger increases in THW after the RtI, and the proposed eHMI C+O was also associated with the lowest likelihood of MV accident occurrence in the AV accident scenario. 
These findings suggest that explicitly externalizing the takeover transition state of a Level~3 AV can help surrounding MV drivers form a more accurate understanding of the AV and promote safer defensive responses during this uncertain transition period, when potential risk is elevated.

\section*{ACKNOWLEDGMENTS}

This work was supported by the Japan Society for the Promotion of Science (JSPS) KAKENHI Grant Numbers 20K19846 and 22H00246, Japan.
The authors used OpenAI ChatGPT (GPT-5.3) for English proofreading and take full responsibility for the final content.

\section*{CRediT author statement}

\textbf{Hailong~Liu}: Conceptualization, Methodology, Software, Validation, Formal Analysis, Visualization, Project administration, Funding acquisition, Writing - Original Draft, Writing - review \& editing.

\textbf{Masaki~Kuge}: Methodology, Software, Investigation, Data Curation, Formal Analysis, Visualization, Writing - review \& editing.

\textbf{Toshihiro~Hiraoka}: Methodology, Funding acquisition, Writing - review \& editing.

\textbf{Takahiro~Wada}: Methodology, Writing - review \& editing.

\footnotesize
\bibliographystyle{IEEEtranN}
\bibliography{main.bib}

\newpage
\begin{IEEEbiography}
[{\includegraphics[width=1in,height=1.25in,clip,keepaspectratio]{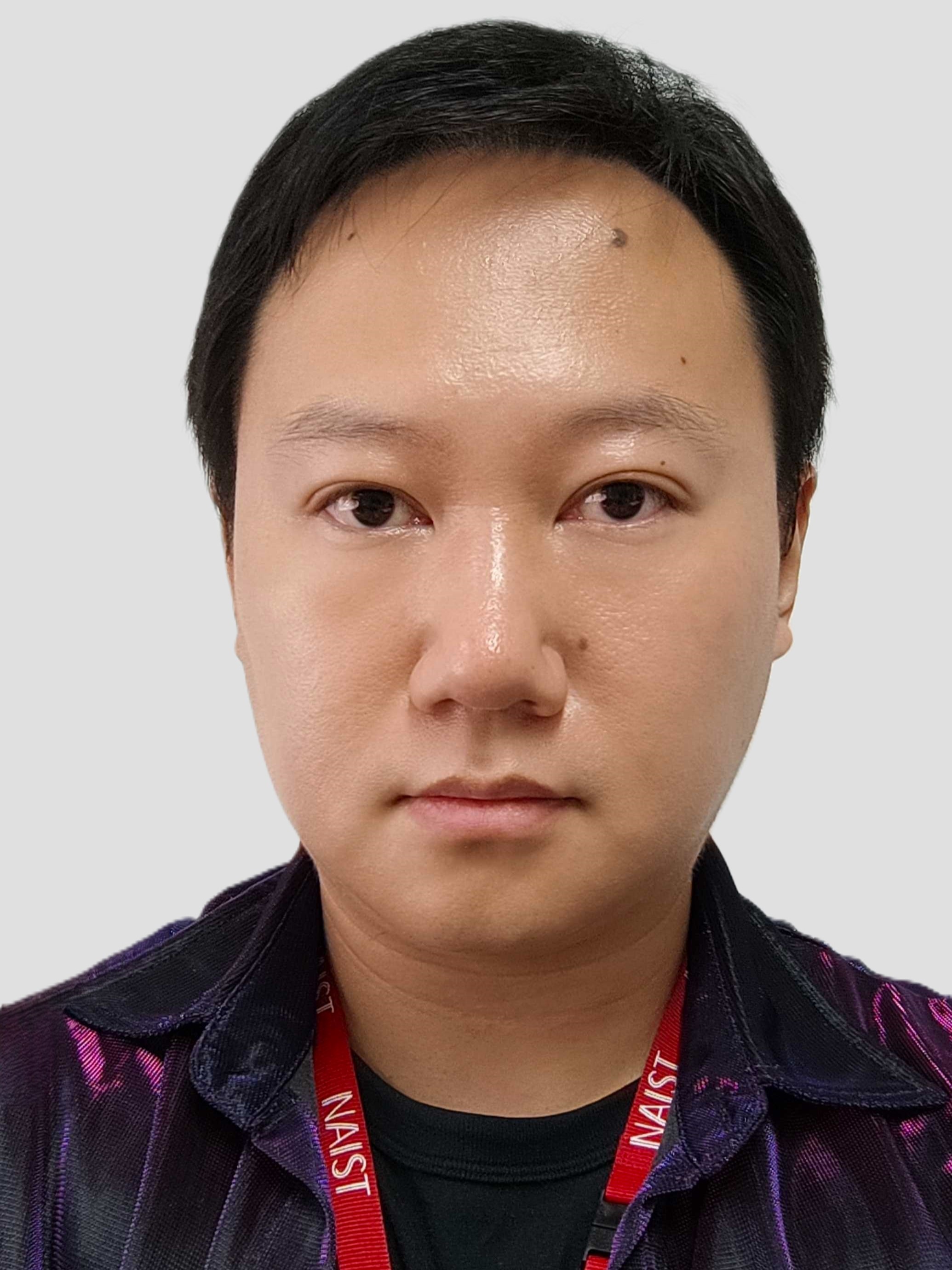}}]
{Hailong~Liu} (S'15--M'19--SM'25) received his B.Eng., M.Eng. and Ph.D. degrees in Engineering from Ritsumeikan University, Japan, in 2013, 2015 and 2018, respectively. He was a JSPS Research Fellow for Young Scientists (DC2) (2016--2018), a researcher at Nagoya University (2018--2021).
In Nov. 2021, he joined Nara Institute of Science and Technology (NAIST), Japan, as an Assistant Professor and was promoted to Associate Professor in Feb. 2024.
From Oct. 2025, he has been a cancer warrior.
His research focuses on human factors and machine learning in intelligent transportation systems. He is a Senior Member of IEEE and holds memberships in IEEE ITSS, RAS, SMC. He also serves on the Human Factors in ITS Committee of IEEE ITSS. In addition, he is a member of JSAE, JSAI, and SICE.
\end{IEEEbiography}

\begin{IEEEbiography}
[{\includegraphics[width=1in,height=1.25in,clip,keepaspectratio,trim=5 0 5 20]{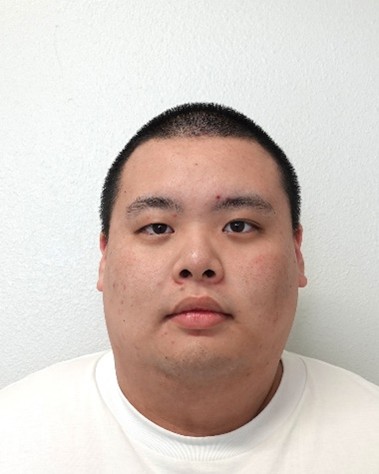}}]
{Masaki~Kuge} received his M.Eng. degree from the Graduate School of Information Science and Engineering, Graduate School of Science and Technology, NAIST, Japan, in 2025.
\end{IEEEbiography}
\vspace{-9mm}

\begin{IEEEbiography}
[{\includegraphics[width=1in,height=1.25in,clip,keepaspectratio,trim=20 0 30 20]{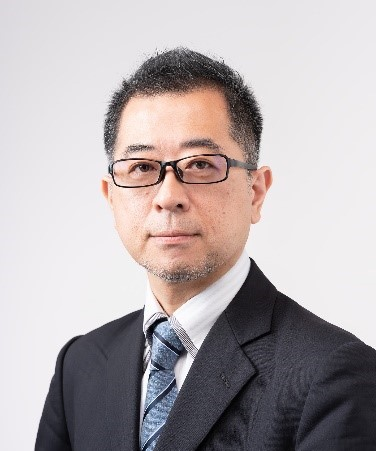}}]
{Toshihiro~Hiraoka} (M'14) received B.E. and M.E. degrees in Precision Engineering in 1994 and 1996, and a Ph.D. in Informatics in 2005, all from Kyoto University, Japan.
He worked at Matsushita Electric Industrial Co., Ltd. (1996-1998), Kyoto University as an Assistant Professor (1998-2017), Nagoya University as a Designated Associate Professor (2017-2019), and the University of Tokyo as a project professor (2019-2022). Since 2022, he has been a Senior Chief Researcher at the Japan Automobile Research Institute. His research interests include human-machine systems, advanced driver-assistance systems, and automated driving systems. He is a member of SICE, HIS, JSAE, JES, IATSS, and IEEE (ITSS).
 \end{IEEEbiography}
\vspace{-9mm}

\begin{IEEEbiography} [{\includegraphics[width=1in,height=1.25in,clip,keepaspectratio]{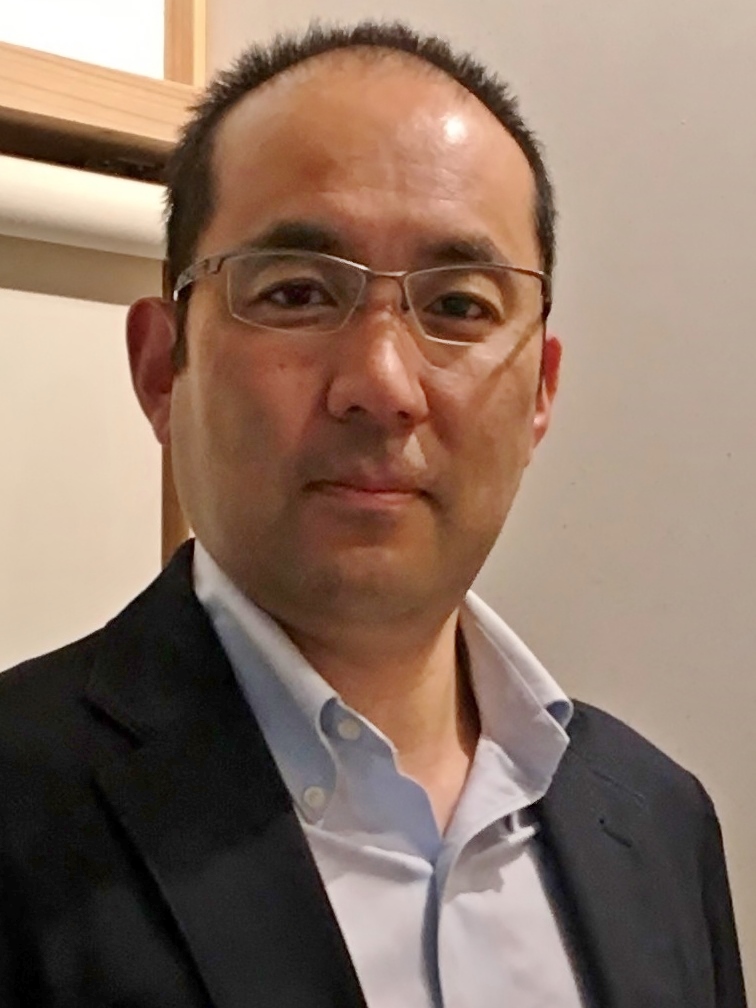}}]{Takahiro~Wada} 
(M'99) received a B.S. degree in Mechanical Engineering, a M.S. degree in Information Science and Systems Engineering, and a Ph.D. degree in Robotics from Ritsumeikan University, Japan, in 1994, 1996, and 1999, respectively. 
He worked as a Research Associate at Ritsumeikan University (1999-2000).
He worked at Kagawa University as a Research Associate (2000-2003), an Assistant Professor (2003--2007) and Associate Professor (2007--2012).  
He has been a full professor at Ritsumeikan University (2012--2021) and at Nara Institute of Science and Technology (2021--present).
His current research interests include robotics, human machine systems, and motion sickness modeling.
He is a member of  IEEE (RAS, ITSS, SMC), SAE, HFES, SICE, JSAE, RSJ, JSME.
\end{IEEEbiography}
\end{document}